\documentclass[12pt]{article}
\usepackage{epsfig}
\usepackage[utf8]{inputenc}
\usepackage[english]{babel}

\begin{document}

\title{
{\bf High-energy elastic diffractive scattering of nucleons in the framework of the two-Reggeon eikonal approximation (from U-70 to LHC)}}
\author{A.A. Godizov\thanks{E-mail: anton.godizov@gmail.com}\\
{\small {\it A.A. Logunov Institute for High Energy Physics}},\\ {\small {\it NRC ``Kurchatov Institute'', 142281 Protvino, Russia}}}
\date{}
\maketitle

\vskip-1.0cm

\begin{abstract}
Elastic diffractive scattering of nucleons is described in terms of Regge-eikonal approach. It is demonstrated that, in a wide kinematic region (starting from the U-70 energies), 
the eikonal of proton-proton scattering can be approximated by the sum of two Reggeon exchange terms (namely, the so-called soft Pomeron and $f$-Reggeon contributions). The range of 
applicability of the considered approximation is determined. The predictive value of the proposed phenomenological scheme is verified.
\end{abstract}

\section*{1. Introduction}

Diffraction phenomena in high-energy hadron physics are related to comparatively large transverse distances between interacting hadrons, from 0.1 to 10 fm. This fact leads to a specific 
situation when the fundamental theory of strong interaction, quantum chromodynamics (QCD), or, more exactly, perturbative QCD, does not help us to describe diffractive scattering of 
hadrons and to make reliable predictions for the observables of the corresponding reactions. Therefore, the most of modern models of hadron diffraction are formulated rather in terms 
of hadrodynamics itself than in terms of interacting quarks and gluons. As a consequence, these models reveal a very low predictive value (for detailed discussion, see a minireview in 
\cite{godizov0}). 

Nonetheless, hadron diffraction processes constitute an important sector of high-energy physics, owing, first of all, to the high fraction of diffraction events in the total number of 
events at high-energy hadron colliders (not less than 30\% at the LHC energies). Hence, approaches are needed which, on the one hand, allow to make reliable predictions (at least, 
qualitative) for hadron diffraction observables, and, on the other hand, admit connection with QCD in the future, in the case of essential progress in nonperturbative QCD methods. 

The aim of this paper is to demonstrate that a simple approximation can be constructed in the framework of the general Regge-eikonal approach, which allows to describe the high-energy 
evolution of the diffraction pattern of nucleon-nucleon elastic diffractive scattering (EDS) within a very large kinematic range. 

The Regge-eikonal formalism itself is expounded in detail in \cite{collins} and \cite{kisselev}. Therefore, in what follows, we restrict ourselves just by a concise review of its main 
points. Next, using some QCD-based arguments, we formulate the two-Reggeon approximation and apply it to available data on high-energy EDS of nucleons, for determination of the 
proposed model applicability range and demonstration of its predictive value. Then, we discuss the issues and compare the considered phenomenological scheme with other models 
of high-energy EDS which exploit the notion of Reggeon. 

\section*{2. Basics of Regge-eikonal approach}

\subsection*{2.1. Eikonal representation and the Van Hove hypothesis}

Regge-eikonal approximation is founded on the eikonal representation of the high-energy elastic scattering amplitude \cite{collins}:
\begin{equation}
\label{eikrepr}
T_N(s,t) = 4\pi s\int_0^{\infty}d(b^2)\,J_0(b\sqrt{-t})\,\frac{e^{2i\delta_N(s,b)}-1}{2i}\;,
\end{equation}
$$
\delta_N(s,b)=\frac{1}{16\pi s}\int_0^{\infty}d(-t)\,J_0(b\sqrt{-t})\,\Omega_N(s,t)\,,
$$
where $s$ and $t$ are the Mandelstam variables, $b$ is the impact parameter, $J_0(x)$ is the Bessel function, $T_N$ is the full amplitude related to strong interaction, the eikonal 
$\delta_N$ is the two-particle-irreducible part of $T_N$ in the coordinate representation, and the function $\Omega_N$ is the eikonal in the transferred momentum representation.

Such a form of the full amplitude resembles the analogous formula for the amplitude of elastic scattering of nonrelativistic spinless particle in external spherically symmetric 
potential field obtained via exact solving of the corresponding Schr\"odinger equation. Then the eikonal is equal to the Born amplitude.

In high-energy hadron physics, the negligibility of multiparticle effects is not evident.\linebreak Nonetheless, we have to presume their subdominance, as proposed in \cite{vanhove}. 
Otherwise, the representation (\ref{eikrepr}) itself does not help us to solve the problem, since it is just reduced to the replacement of the unknown function of two variables, 
$T_N(s,t)$, by another unknown function, $\Omega_N(s,t)$, without any specification of the functional form of $\Omega_N(s,t)$.

Thus, the Van Hove hypothesis that, at high energies and low tranferred momenta, the $s$-channel eikonal for elastic scattering of two spinless hadrons can be represented as 
\begin{equation}
\label{eik0}
\Omega_N(s,t)=\sum_j\sum_{m_j}\frac{\xi_jB^{(j,m_j)}(s,t)}{m_j^2-t}+...
\end{equation}
is the first basic assumption necessary for construction of the Regge-eikonal approximation for the amplitude $T_N(s,t)$. Here $\xi_j$ is the signature factor, 
$B^{(j,m_j)}(s,t)\equiv\beta^{(j,m_j)}(t)P_j(z_t)$, $P_j(x)$ is the Legendre polynomial of degree $j$, $z_t$ is the cosine of the scattering angle in the $t$-channel, 
$m_j^2\equiv M_j^2-i M_j\Gamma_j$, where $j$, $M_j$, and $\Gamma_j$ are the resonance spin, mass, and full width of decay, correspondingly, and ``...'' denotes subdominant nonresonance 
contributions. 

\subsection*{2.2. Factorization of the residues}

However, the approximation (\ref{eik0}) for the eikonal does not imply, by default, any factorization of the functions $\beta^{(j,m_j)}(t)$ into two factors each related directly to a 
certain one of the colliding spinless hadrons. Such a factorization emerges if we follow the formalism developed in \cite{kisselev}, wherein the residue $B^{(j,m_j)}(s,t)$ is 
represented as the convolution of two currents dependent on the 4-momenta $p_1$ or $p_2$ of the interacting hadrons and on the transferred 4-momentum $\Delta$ ($t\equiv\Delta^2$) with 
some tensor related to the exchanged virtual resonance and dependent on $\Delta$ only:
\begin{equation}
\label{res1}
B^{(j,m_j)}(s,t) = J^{(h_1,m_j)}_{\mu_1...\mu_j}(p_1,\Delta)\;D_{(m_j)}^{\mu_1...\mu_j;\,\nu_1...\nu_j}(\Delta)\;J^{(h_2,m_j)}_{\nu_1...\nu_j}(p_2,-\Delta)\,.
\end{equation}
Here $J^{(h_{1,2},m_j)}_{\mu_1...\mu_j}(p_{1,2},\Delta)$ are the currents of the colliding particles, the indices $h_{1,2}$ denote the sort of the hadron, and the index $m_j$ indicates 
that the properties of $J$ and $D$ depend on the physical structure of the exchanged virtual resonance state (below it will be omitted).

If we represent the tensor structure of $J^{(h)}_{\mu_1...\mu_j}(p,\Delta)$ in the most general form, 
\begin{equation}
\label{str1}
J^{(h)}_{\mu_1...\mu_j}(p,\Delta)=g^{(h)}_j(t)\,p_{\mu_1}...\,p_{\mu_j}\,s_0^{-j/2}\;+\;...
\end{equation}
(where $s_0 = 1$ GeV$^2$ is the unit of measurement, while ``...'' denotes the combined contribution of the tensor substructures different from $p_{\mu_1}...\,p_{\mu_j}$) and convolve 
both the currents with 
\begin{equation}
\label{str2}
D^{\mu_1...\mu_j;\,\nu_1...\nu_j}(\Delta)\;=\;\frac{f_j(t)}{j!}\,[\,g^{\mu_1\nu_1}...\,g^{\mu_j\nu_j}\;\;+\;\;all\;the\;permutations\;of\;\nu_k]\;+\;...\;,
\end{equation}
then we obtain
\begin{equation}
\label{ome1}
B^{(j)}(s,t) = g^{(h_1)}_j(t)\,g^{(h_2)}_j(t)\,f_j(t)\left(\frac{(p_1p_2)}{s_0}\right)^j\left[1\,+\,O\left((p_1p_2)^{-1}\right)\right].
\end{equation}
Hence, in the kinematic region $(p_1+p_2)^2\equiv s\gg \{p_{1,2}^2,s_0,|t|\}$, we come to the following high-energy approximation for the residue $B$:
\begin{equation}
\label{ome2}
B^{(j)}(s,t)\approx\tilde g^{(h_1)}_j(t)\,\tilde g^{(h_2)}_j(t)\left(\frac{s}{2s_0}\right)^j,
\end{equation}
where the factors $\tilde g^{(h_{1,2})}_j(t)\equiv g^{(h_{1,2})}_j(t)\,f^{1/2}_j(t)$ could be interpreted as the effective couplings of the exchanged virtual resonance to the 
interacting hadrons.

\subsection*{2.3. Helicity currents of protons}

In the case of the high-energy scattering of protons, we should not {\it a priori} ignore the spin-flip effects, but it is possible to show that they are subdominant at small enough 
values of the transferred momentum.

Let $u^{(1)}(p)$ ($u^{(2)}(p)$) be the Dirac spinor related to the proton state with 4-momentum $p$ and positive (negative) helicity, and let the normalization of these spinors be fixed 
by the relation 
\begin{equation}
\label{norm}
\sum_i u^{(i)}_\alpha(p)\,\bar u^{(i)}_\beta(p) = (\hat p + m_p\, I)_{\alpha\beta}\,,
\end{equation}
where $\alpha$ and $\beta$ are the spinor indices, $m_p$ is the proton mass, $I$ is the unity matrix, $\bar u^{(i)}\equiv u^{(i)+}\gamma^0$, $\hat p\equiv p_\mu \gamma^\mu$, and 
$\gamma^\mu$ are the Dirac matrices. Then, the most general form for an arbitrary helicity current of proton is 
\begin{equation}
\label{heli}
J_{(i\,i')}^{\mu_1...\mu_j}(p,\Delta)=\bar u^{(i')}_\alpha(p+\Delta)\left(g^{(0)}_j(t)\,I_{\alpha\beta}\,p^{\mu_1}...\,p^{\mu_j}\,s_0^{-j/2}+\right.
\end{equation}
$$
\left.+\;\frac{g^{(1)}_j(t)}{j}\left[\gamma^{\mu_1}_{\alpha\beta}\,p^{\mu_2}...\,p^{\mu_j}\;+\;permutations\;between\;\mu_1\;and\;\mu_k\right]s_0^{-(j-1)/2}
\;+\;...\right)u^{(i)}_\beta(p)\,.
$$
Taking into account that $s\gg \{m_p^2,s_0,|t|\}$ and summing over the spinor indices, we obtain, in the leading approximation, that 
\begin{equation}
\label{flip}
J_{2\,1}^{\mu_1...\mu_j}(p,\Delta)=-J_{1\,2}^{\mu_1...\mu_j}(p,\Delta)=\sqrt{-t}\,g^{(0)}_j(t)\,p^{\mu_1}...\,p^{\mu_j}\,s_0^{-j/2}\,+\,...
\end{equation}
and
\begin{equation}
\label{nonflip}
J_{1\,1}^{\mu_1...\mu_j}(p,\Delta)=J_{2\,2}^{\mu_1...\mu_j}(p,\Delta)=\left[2m_p\,g^{(0)}_j(t)\,+\,2\sqrt{s_0}\,g^{(1)}_j(t)\right]p^{\mu_1}...\,p^{\mu_j}\,s_0^{-j/2}\,+\,...
\end{equation}

Consequently, we can ignore the spin effects in the region of small enough values of the transferred momentum, wherein the following inequality is valid:
\begin{equation}
\label{suppr}
\sqrt{-t}\,|g^{(0)}_j(t)|\ll |2m_p\,g^{(0)}_j(t)\,+\,2\sqrt{s_0}\,g^{(1)}_j(t)|\equiv|g^{(p)}_j(t)|\,.
\end{equation}

The measurements performed by STAR Collaboration \cite{star1} demonstrate that, in the range $\sqrt{-t}<0.2$ GeV, the spin-flip events take place due to electromagnetic interaction 
only. In their turn, available data on the proton-proton EDS at lower energies \cite{pol1,pol2,pol3} point to the fact that, in the interval 0.5 GeV $<\sqrt{-t}<$ 1.5 GeV, the 
spin-flip contributions affect the differential cross-section very slightly, and, thus, they could be considered negligible, in the leading approximation (detailed discussion of this 
matter can be found in a recent paper \cite{selyugin1}). Therefore, in further, we assume their subdominance in the whole kinematic range relevant for the high-energy EDS of nucleons. 
In this case, we may treat nucleons as spinless particles, and direct exploitation of the Regge-eikonal representation (\ref{eikrepr}) of the scattering amplitude is justified.

\subsection*{2.4. Reggeization}

Next, let us consider the combined contribution into the elastic scattering eikonal by a family of $t$-channel resonance terms, and let every term in this family to be related to some 
meson or glueball state with nonzero even value of the spin $j$ ($j = 2,4,6,...$). As it has been discussed above, in the kinematic sector $s\gg \{p_{1,2}^2,s_0,|t|\}$ of the 
$s$-channel physical region, this contribution can be approximated by the expression 
\begin{equation}
\label{regge1}
\Omega_+(s,t) = \sum_{j=2}^{\infty}\frac{1 + e^{-i\pi j}}{m_j^2-t}\left(\frac{s}{2s_0}\right)^j\tilde g_j^{(h_1)}(t)\,\tilde g_j^{(h_2)}(t)\,.
\end{equation}
If $\tilde g_j^{(h)}(t)$ and $m_j^2$ are the values at positive even $j$ of some analytic functions which are holomorphic in the region ${\rm Re}\,j > 0$ and behave as $O(e^{k|j|})$, 
$k<\pi$, at $j\to\infty$, then, under the Carlson theorem, the unilocal analytic continuation of (\ref{regge1}) is possible into the region of complex $j$ (the Regge hypothesis 
\cite{collins}). We denote these functions by $\tilde g^{(h)}(t,j)$ and $m^2(j)$, respectively. Via the Sommerfeld-Watson transform \cite{collins,sommer}, we can replace the sum over 
$j$ in (\ref{regge1}) by the integral over the contour $C$ encircling the real positive half-axis in the complex $j$-plane, including the point $j = 2$ (but not including the point 
$j = 0$), in such a way that the half-axis is on the right:
\begin{equation}
\label{regge2}
\Omega_+(s,t) = -\frac{1}{2i}\oint_C\frac{dj}{\sin(\pi j)}\;\frac{1+e^{-i\pi j}}{m^2(j)-t}\left(\frac{s}{2s_0}\right)^j\tilde g^{(h_1)}(t,j)\,\tilde g^{(h_2)}(t,j)\,.
\end{equation}

According to our assumptions, the only sources of the integrand singularities in the region ${\rm Re}\,j > 0$ are the zeros of the functions $\sin(\pi j)$ and $m^2(j)-t$. Hence, 
deforming the contour $C$ to the axis ${\rm Re}\,j = \epsilon\to +0$ (the asymptotic behavior of the functions $\tilde g^{(h)}(t,j)$ at $j\to\infty$ in the region ${\rm Re}\,j >0$ and 
$t\le 0$ should allow such a deformation), we obtain 
\begin{equation}
\label{regge3}
\Omega_+(s,t) = \; - \; \frac{1}{2i}\int_{\epsilon-i\infty}^{\epsilon+i\infty}\frac{dj}{\sin(\pi j)}\;\frac{1+e^{-i\pi j}}{m^2(j)-t}
\left(\frac{s}{2s_0}\right)^j\tilde g^{(h_1)}(t,j)\,\tilde g^{(h_2)}(t,j) \; - 
$$
$$
- \; \frac{1 + e^{-i\pi\alpha(t)}}{\sin(\pi\alpha(t))}\;\pi\,\alpha'(t)\left(\frac{s}{2s_0}\right)^{\alpha(t)}g^{(h_1)}(t)\,g^{(h_2)}(t)\,.
\end{equation}
Here $g^{(h)}(t)\equiv\tilde g^{(h)}(t,\alpha(t))$, while the analytic function $\alpha(t)$ is the root of the equation\linebreak $m^2(j)\;-\;t = 0$. We assume, for simplicity, that, at 
$t\le 0$, this equation has a unique $t$-dependent root in the region ${\rm Re}\,j > 0$, wherein the functions $\tilde g^{(h)}(t,j)$ and $m^2(j)$ are holomorphic. Thus, $\alpha(t)$ is 
related to a moving pole in the complex $j$-plane. Such poles are called Regge poles, and the corresponding functions $\alpha(t)$ are called Regge trajectories. 

As far as ${\rm Re}\,\alpha(t)>0$, the background integral is negligible at high enough values of $s$, and, finally, we come to the following high-energy approximation for the 
contribution of any even-spin Reggeon to the eikonal:
\begin{equation}
\label{regge4}
\Omega_+(s,t)\approx\xi_+(\alpha(t))\;\alpha'(t)\left(\frac{s}{2s_0}\right)^{\alpha(t)}g^{(h_1)}(t)\,g^{(h_2)}(t)\,,
\end{equation}
where $\xi_+(\alpha(t))$ denotes the so-called Reggeon signature factor for the even-spin Reggeon associated with the Regge trajectory $\alpha(t)$: 
\begin{equation}
\label{sign1}
\xi_+(\alpha(t))=\pi\left(i+{\rm tan}\frac{\pi(\alpha(t)-1)}{2}\right)\,.
\end{equation}
In the case of an odd-spin Reggeon, the corresponding factor would be \cite{collins} 
\begin{equation}
\label{sign2}
\xi_-(\alpha(t))=\pi\left(i-{\rm cot}\frac{\pi(\alpha(t)-1)}{2}\right)\,.
\end{equation}

Formula (\ref{regge4}) together with the eikonal representation (\ref{eikrepr}) of the elastic scattering amplitude, where the eikonal is just the sum of a few Reggeon contributions, 
is the essence of Regge-eikonal approach.

\section*{3. Two-Reggeon eikonal model}

\subsection*{3.1. Specific structure of the eikonal}

\begin{figure}[ht]
\begin{center}
\epsfxsize=8cm\epsfysize=8cm\epsffile{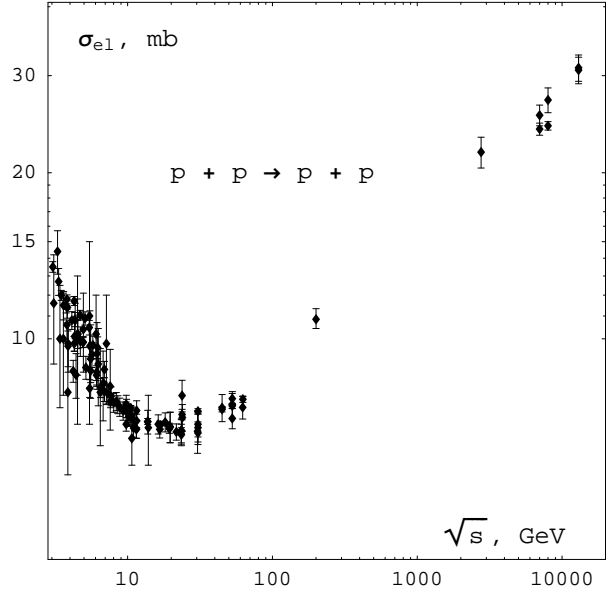}
\end{center}
\vskip -0.5cm
\caption{The energy evolution of the $p\,p$ EDS integrated cross-section \cite{ppel,star2}.}
\label{elapp}
\end{figure}

The main problem emerging in the framework of Regge-eikonal approach and, in general, in Regge theory is the evident functional indeterminacy. In particular, describing the 
high-energy EDS of nucleons, we have to introduce two unknown functions for every Reggeon term, namely, the Regge trajectory and the corresponding Reggeon coupling to nucleon. The 
energy evolution of the integrated cross-section of proton-proton elastic scattering \cite{ppel,star2} (see Fig. \ref{elapp}) points to the existence of two different sorts of 
Reggeons:
\begin{itemize}
\item supercritical Reggeons with the Regge trajectory intercept higher than the unity which are responsible for the cross-section growth at high energies;
\item secondary Reggeons responsible for the cross-section decrease at low energies. 
\end{itemize}
In literature, secondary Reggeons are usually associated with the families of the Quark Model low-mass vector and tensor mesons ($\omega$, $f$, $\rho$, $a$, $\phi$, {\it etc.}), while 
the even-spin (odd-spin) supercritical Reggeons called Pomerons (Odderons) are assumed to be related to glueballs.

\begin{figure}[ht]
\hskip 0.5cm
\epsfxsize=6.7cm\epsfysize=4.7cm\epsffile{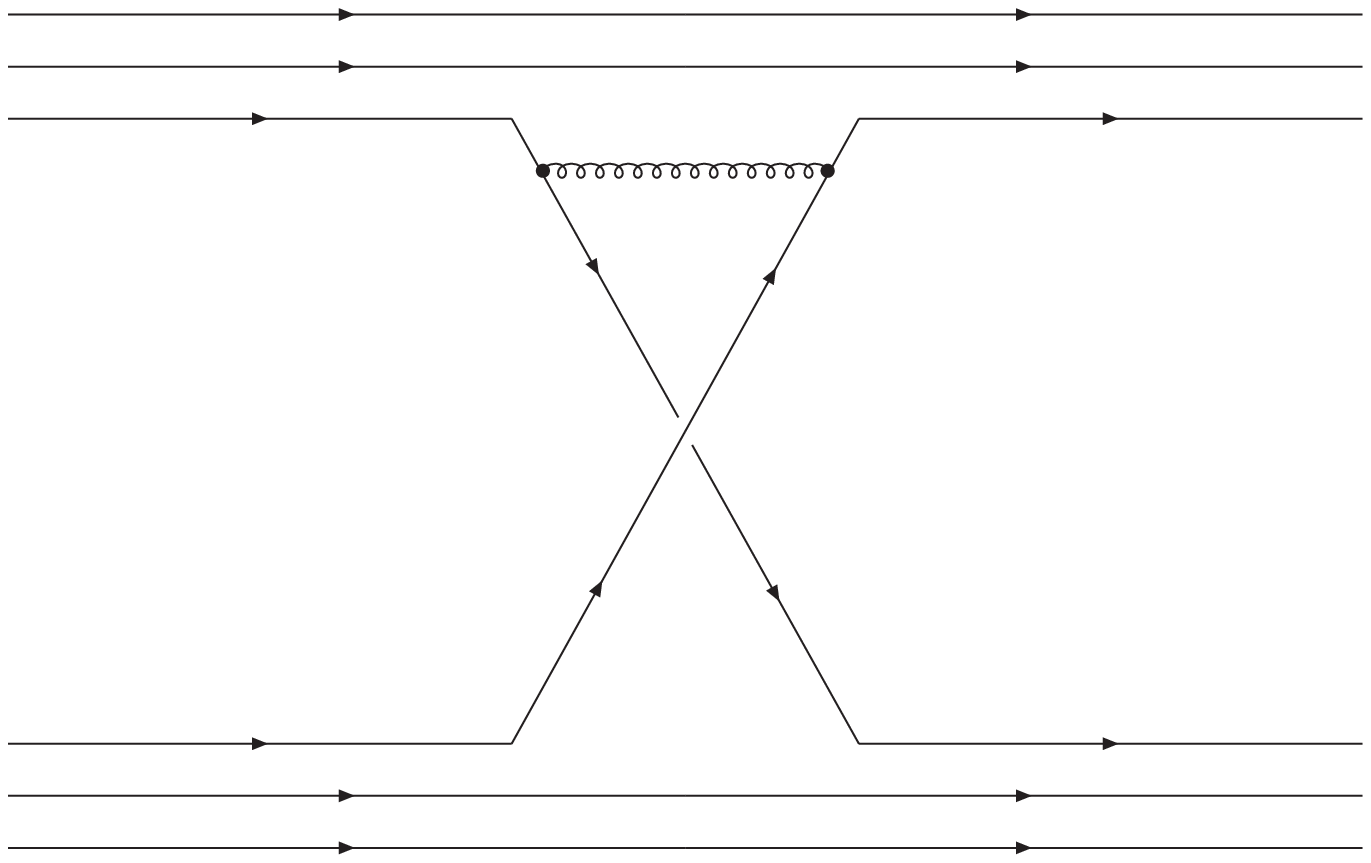}
\vskip -4.75cm
\hskip 9.8cm
\epsfxsize=6.7cm\epsfysize=4.7cm\epsffile{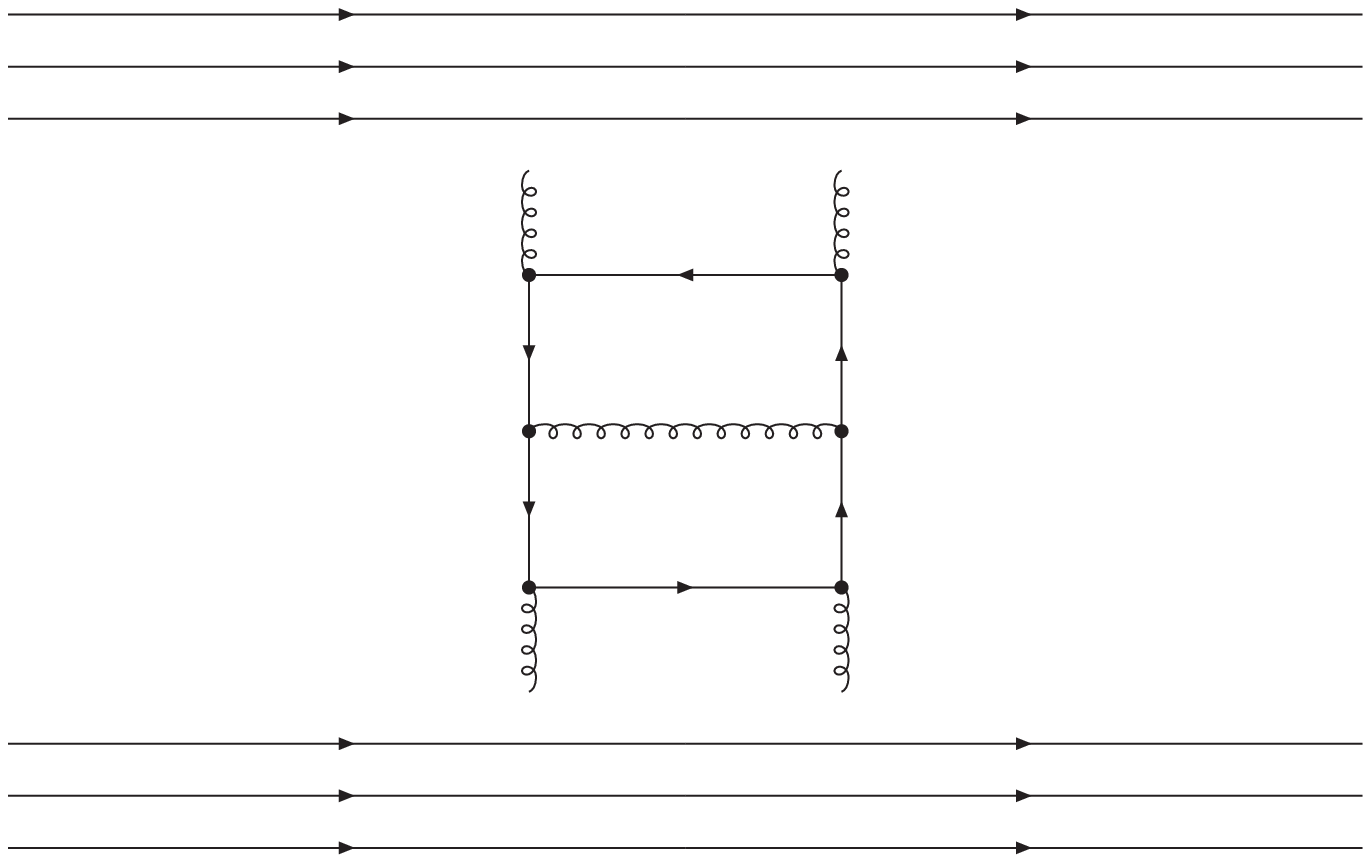}
\caption{Two types of the QCD contributions to nucleon-nucleon EDS wherein the meson resonance exchange terms emerge.}
\label{qcd}
\end{figure}

The most important question is how many Reggeons contribute crucially into the eikonal of nucleon-nucleon EDS at available energies. Different models give different answers. 
However, the fundamental quantum field model of strong interaction helps to clarify the general structure of the secondary Reggeon part of the nucleon-nucleon EDS eikonal. 
Namely, two types of the secondary Reggeon exchange contributions emerge in the QCD dynamics of nucleons:
\begin{itemize}
\item the contributions of type 1 related to the exchange by valence quarks between nucleons (the left picture in Fig. \ref{qcd});
\item the contributions of type 2 related to those QCD terms wherein the exchange by valence quarks does not take place (the right picture in Fig. \ref{qcd}). 
\end{itemize}
\vskip -0.5cm
\vskip -0.3cm

Moreover, it was explicitly shown in the framework of the so-called dual approximation of QCD \cite{rossi} that the contributions of type 1 to nucleon-nucleon interaction are 
purely real. Therefore, if we represent the secondary Reggeon part of the nucleon-nucleon EDS eikonal in terms of $C$-even and $C$-odd Reggeon exchanges, then we come to the following 
structure of the nucleon-nucleon and nucleon-antinucleon diffractive interaction:
\begin{equation}
\label{eiksec}
\Omega_N^{(sec)}(s,t) \approx \Omega_{\rm FR}(s,t)+\Omega_f(s,t)\mp\Omega_\omega(s,t)+\Omega_a(s,t)\mp\Omega_\rho(s,t)\,,
\end{equation}
where $\Omega_{\rm FR}(s,t)$ is the $f$-Reggeon (FR) contribution of type 2 related to the QCD terms without exchange by valence quarks between the nucleons, 
$\Omega_f(s,t)$ is the contribution of type 1 of the same Reggeon, the last three terms are related to the secondary Reggeons $\omega$, $a$, and $\rho$, and the sign ``$-$'' (``$+$'') 
at the $C$-odd terms corresponds to the nucleon-nucleon (nucleon-antinucleon) interaction.

If the imaginary parts of $\Omega_f(s,t)$ and $\Omega_\omega(s,t)$ and, also, of $\Omega_a(s,t)$ and $\Omega_\rho(s,t)$ coincide (the so-called exchange degeneracy phenomenon), then 
they annihilate each other in the $p\,p$ EDS regime, in accordance with the physical pattern emerging in the framework of the dual approximation of QCD. Therefore, in the case of the 
$p\,p$ EDS, the contributions of type 1 can be neglected, in the leading approximation, and, thus, only the FR exchange term of type 2 is crucial for the $p\,p$ EDS.\footnote{The only 
exclusion is a very small vicinity of the diffraction dip in the diffraction pattern of the $p\,p$ EDS at low enough values of the collision energy, wherein the combined secondary 
Reggeon contribution into the eikonal real part has an essential impact on the depth and shape of the diffraction dip.}

Regarding the supercritical Reggeons, the situation is much more indefinite. No argument exists which limits {\it a priori} the number of crucial Pomerons or Odderons. As a consequence, 
their amount varies from model to model. In this paper, we restrict ourselves by the simplest variant, wherein just one supercritical Reggeon is assumed to give the leading contribution 
within the relevant kinematic range, namely, the so-called soft Pomeron (SP). Thus, the considered approximation is reduced to the representation of the $p\,p$ EDS eikonal as the 
sum of two Regge-pole terms related to the SP and the FR:
\begin{equation}
\label{eik2}
\Omega_N(s,t) \approx \Omega_{\rm SP}(s,t) + \Omega_{\rm FR}(s,t)\,.
\end{equation}

Below, it will be demonstrated that this simplified phenomenological scheme (which ignores the secondary Reggeon contributions of type 1) allows to reach a relatively good description 
of available data on the high-energy EDS of protons. As well, we will determine the kinematic range of applicability of approximation (\ref{eik2}) and check its predictive value via 
application to available data at ultrahigh energies.

\subsection*{3.2. General properties of the Regge trajectories and the Reggeon couplings to proton}

Hereby, in the framework of the proposed approximation, we have four unknown functions, namely, the Regge trajectories of the SP and the FR, and their couplings to proton: 
$\alpha_{\rm SP}(t)$, $\alpha_{\rm FR}(t)$, $g^{(p)}_{\rm SP}(t)$, and $g^{(p)}_{\rm FR}(t)$. The main property of these functions is that they take real values at real values of $t$ 
below the two-pion threshold \cite{collins}: 
\begin{equation}
\label{proper}
{\rm Im}\,\alpha_{\rm SP}(t)=0\,,\;\;{\rm Im}\,\alpha_{\rm FR}(t)=0\,,\;\;{\rm Im}\,g^{(p)}_{\rm SP}(t)=0\,,\;\;{\rm Im}\,g^{(p)}_{\rm FR}(t)=0\;\;\;\;(-\infty<t<4m_{\pi^0}^2)\,.
\end{equation}
The absence of any points in $\alpha_{\rm SP}(t)$ and $\alpha_{\rm FR}(t)$ which are related to scalar mesons imposes the following restriction on their intercept values:
\begin{equation}
\label{proper2}
\alpha_{\rm SP}(0)>0\,,\;\;\alpha_{\rm FR}(0)>0\,.
\end{equation}
The unitarity condition for the elastic scattering amplitude in the impact parameter representation \cite{collins},
\begin{equation}
\label{unitar}
0\le |T_N(s,b)|^2\le {\rm Im}\,T_N(s,b)\le 1\,,
\end{equation}
reguires the eikonal imaginary part to be positive:
\begin{equation}
\label{unitar2}
{\rm Im}\,\delta_N(s,b)\ge 0\,.
\end{equation}
In its turn, the latter inequality takes place only if the functions $\alpha_{\rm SP}(t)$ and $\alpha_{\rm FR}(t)$ grow monomonically in the region of negative $t$:
\begin{equation}
\label{proper3}
\alpha'_{\rm SP}(t)>0\,,\;\;\alpha'_{\rm FR}(t)>0\;\;\;\;(-\infty<t\le 0)\,.
\end{equation}

If the SP is related to some family of even-spin glueballs, and the quark-antiquark fraction in the SP content is negligible, then some arguments emerge \cite{kearney} that the SP 
Regge trajectory tends to the unity in the limit of asymptotically high transferred momenta:
\begin{equation}
\label{asypom}
\lim_{t\to -\infty}\alpha_{\rm SP}(t) = 1\,.
\end{equation}
The analogous limit for the FR Regge trajectory is \cite{kwiecinski}
\begin{equation}
\label{asyfre}
\lim_{t\to -\infty}\alpha_{\rm FR}(t) = 0\,.
\end{equation}
Thus, at $t\to -\infty$, the SP (FR) can be treated as a vector (scalar) virtual object. Hence, according to the so-called quark counting rules \cite{matveev}, the asymptotic behavior 
of the SP and FR couplings to proton is expected to be 
\begin{equation}
\label{asypom2}
\lim_{t\to -\infty}g^{(p)}_{\rm SP}(t) = O(|t|^{-2})\,,\;\;\lim_{t\to -\infty}g^{(p)}_{\rm FR}(t) = O(|t|^{-3/2})\,.
\end{equation}

\vskip -0.4cm
\begin{figure}[ht]
\epsfxsize=8.2cm\epsfysize=8.2cm\epsffile{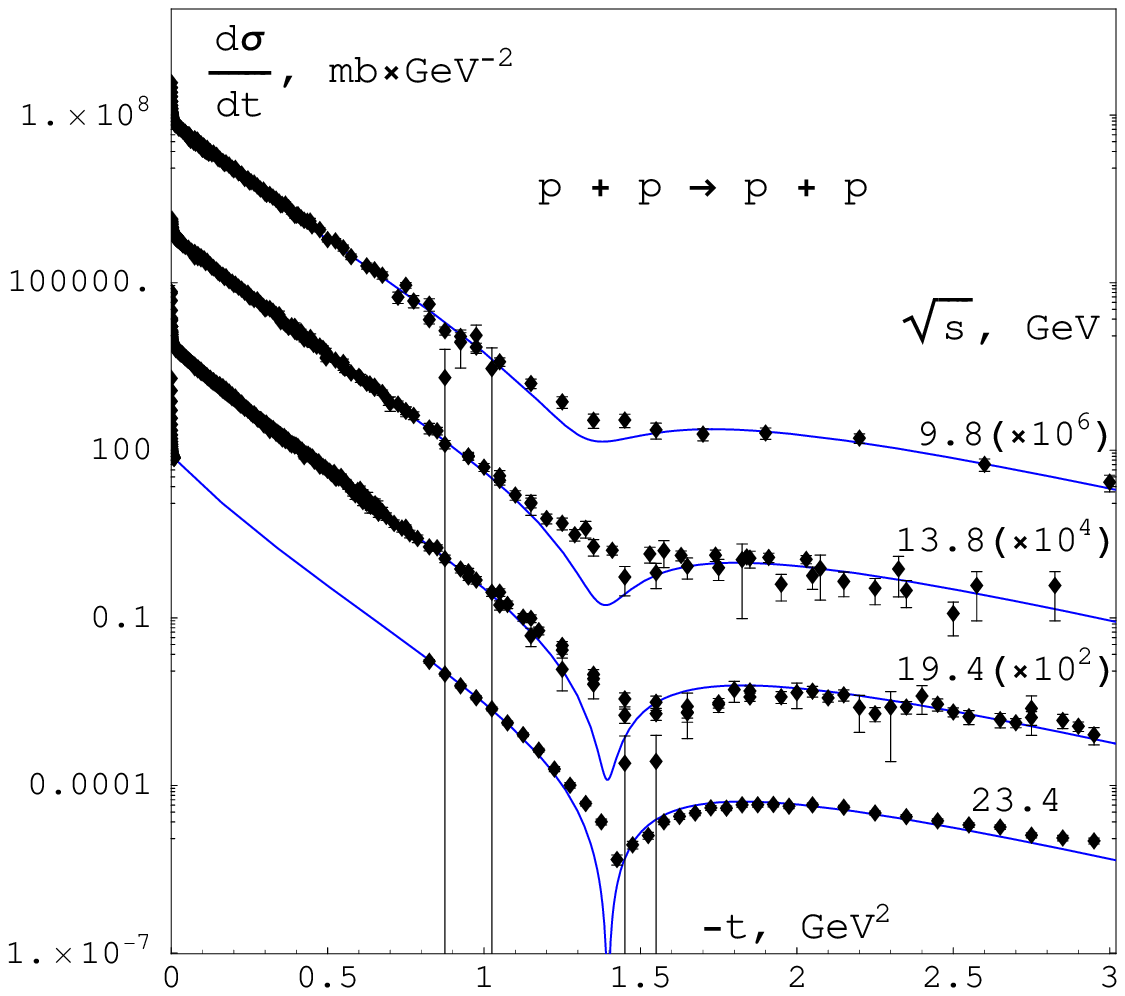}
\vskip -8.25cm
\hskip 8.8cm
\epsfxsize=8.2cm\epsfysize=8.2cm\epsffile{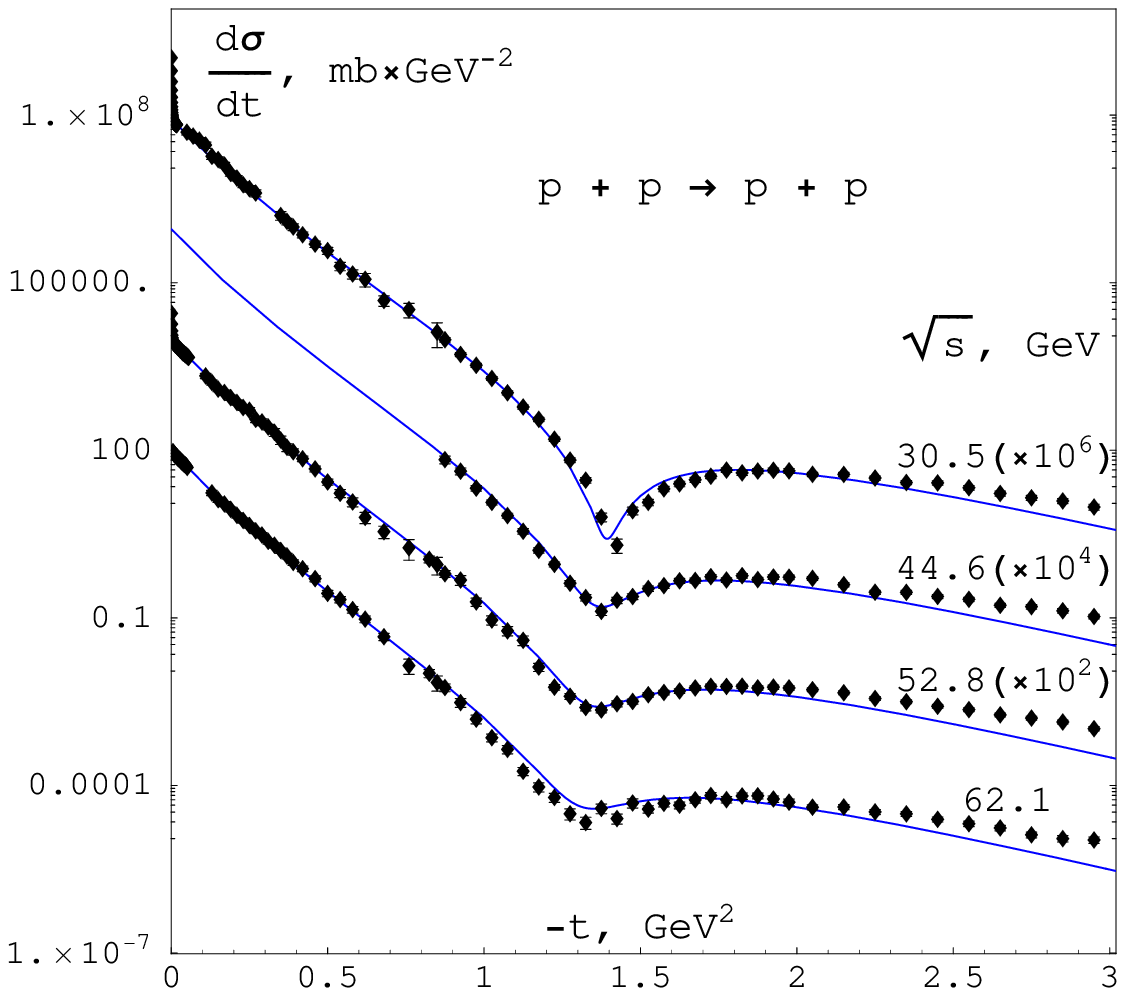}
\caption{The differential cross-sections of proton-proton EDS at high energies and low transferred momenta.}
\label{pplow}
\end{figure}

\subsection*{3.3. Test parametrizations for the unknown functions}

Unfortunately, the above-mentioned restrictions on the $t$-behavior of the Regge trajectories and the Reggeon couplings to proton do not allow to determine unambiguously 
their analytic form. Therefore, in order to apply the proposed model to available experimental data, we have to introduce some test parametrizations. The simplest functions which 
satisfy the relations (\ref{proper}), (\ref{proper2}), and (\ref{proper3}) -- (\ref{asypom2}) are the following ones:
\begin{equation}
\label{pomeron}
\alpha_{\rm SP}(t) = 1+\frac{\alpha_{\rm SP}(0)-1}{1-\frac{t}{\tau_{\rm SP}}}\;,\;\;\;\;g^{(p)}_{\rm SP}(t)=\frac{g^{(p)}_{\rm SP}(0)}{(1-a_1\,t)^2}\;,
\end{equation}
$$
\alpha_{\rm FR}(t) = \frac{\alpha_{\rm FR}(0)}{1-\frac{t}{\tau_{\rm FR}}}\;,\;\;\;\;g^{(p)}_{\rm FR}(t)=\frac{g^{(p)}_{\rm FR}(0)}{(1-a_2\,t)^{3/2}}\;.
$$
These parametrizations should be considered just as some rough phenomenological approximations to $\alpha_{\rm SP}(t)$, $g^{(p)}_{\rm SP}(t)$, $\alpha_{\rm FR}(t)$, and 
$g^{(p)}_{\rm FR}(t)$ in the region of negative $t$ only. The true Regge trajectories and Reggeon couplings to proton have much more complicated $t$-behavior with branching 
points, and, certainly, they have no poles on the physical sheet. Therefore, the approximations (\ref{pomeron}) are absolutely invalid in the region of positive values of $t$, 
because the behavior $\alpha(t)\to\infty$ at finite $t$-values is impossible for true Regge trajectories.

Having substituted the functions (\ref{pomeron}) into (\ref{regge4}), we obtain the expressions for the corresponding Reggeon contributions to the eikonal. Then, with the help of 
the approximation (\ref{eik2}) and the eikonal representation (\ref{eikrepr}) of the elastic scattering amplitude, we can treat the differential cross-section,
\begin{equation}
\label{diffsech}
\frac{d\sigma_{el}}{dt} = \frac{|T_N(s,t)|^2}{16\pi s^2}\,,
\end{equation}
as a function of $s$ and $t$ dependent, as well, on the model free parameters.

The next step is the fitting of the free parameter values to available data on the proton-proton EDS in the relevant kinematic range.

\section*{4. Applying the model to experimental data on the\linebreak nucleon-nucleon EDS angular distributions}

As we intend not only to describe available data, but, also, to check the model predictive value, so we should, first, fit the model parameter values to the limited dataset in the 
restricted kinematic range and, second, compare the model predictions for the differential cross-sections beyond this range with those data which were not included into the fitting 
procedure. For this purpose, the following set of data on the proton-proton EDS is chosen: the ISR data \cite{amos,nagy} in the range \{30 GeV $<\sqrt{s}<$ 63 GeV, 
0.1 GeV $\le\sqrt{-t}\le$ 1.5 GeV\}, the high precision data at $\sqrt{s}=$ 9.8 GeV and 0.01 GeV$^2\le -t\le$ 0.75 GeV$^2$ gathered at U-70 \cite{geshkov} and the Fermilab 
accellerator \cite{ayres}, and the data at $\sqrt{s}=$ 200 GeV and 0.045 GeV$^2\le -t\le$ 0.135 GeV$^2$ recently published by STAR Collaboration \cite{star2}. 

The fitting results are presented in Tables \ref{tab1} and \ref{tab2} and in Fig. \ref{pplow}. As well, Fig. \ref{pplow} contains the ISR data in the range \{30 GeV $<\sqrt{s}<$ 63 GeV, 
0.05 GeV$^2\le -t\le$ 0.85 GeV$^2$\} \cite{break0}, the data at\linebreak $\sqrt{s}=$ 23.4 GeV \cite{amos,nagy}, and the data at lower energies taken from the collection \cite{cudell}. 
We emphasize once again that only those data were included into the fitting procedure which were mentioned in Table \ref{tab2}. The model curves outside of the kinematic ranges pointed 
out in this table should be considered as the predictions.

\begin{table}[ht]
\begin{center}
\begin{tabular}{|l|l|}
\hline
\bf Parameter          & \bf Value        \\
\hline
$\alpha_{\rm SP}(0)-1$ & 0.114            \\
$\tau_{\rm SP}$        & 0.552 GeV$^2$    \\
$g^{(p)}_{\rm SP}(0)$  & 13.1 GeV         \\
$a_1$                  & 0.276 GeV$^{-2}$ \\
$\alpha_{\rm FR}(0)$   & 0.61             \\
$\tau_{\rm FR}$        & 1.54 GeV$^2$     \\
$g^{(p)}_{\rm FR}(0)$  & 18.2 GeV         \\
$a_2$                  & 0.47 GeV$^{-2}$  \\
\hline
\end{tabular}
\end{center}
\caption{The parameter values for (\ref{pomeron}) obtained via fitting to the limited dataset in the kinematic range \{9.8 GeV $\le\sqrt{s}\le$ 200 GeV, 
0.1 GeV $\le\sqrt{-t}\le$ 1.5 GeV\}.}
\label{tab1}
\end{table}

\begin{table}[ht]
\begin{center}
\begin{tabular}{|l|l|l|l|l|}
\hline
\bf Reference & $\sqrt{s}$, GeV & $t$-{\bf interval} & \bf Number of points & \bf $\chi^2$ \\
\hline
 \cite{geshkov} &   9.8  &  0.0124 GeV$^2$ $\le -t\le$ 0.12  GeV$^2$ & 14  &  42  \\
 \cite{ayres}   &   9.8  &  0.0375 GeV$^2$ $\le -t\le$ 0.75  GeV$^2$ & 16  &  38  \\
 \cite{amos}    &  30.7  &  0.01   GeV$^2$ $<-t<$ 0.018 GeV$^2$      & 10  &   3  \\
 \cite{nagy}    &  30.5  &  0.875  GeV$^2$ $\le -t\le$ 2.25  GeV$^2$ & 26  & 136  \\
 \cite{nagy}    &  44.6  &  0.875  GeV$^2$ $\le -t\le$ 2.25  GeV$^2$ & 26  & 105  \\
 \cite{amos}    &  52.8  &  0.01   GeV$^2$ $<-t<$ 0.056 GeV$^2$      & 23  &  57  \\
 \cite{nagy}    &  52.8  &  0.825  GeV$^2$ $\le -t\le$ 2.25  GeV$^2$ & 27  & 157  \\
 \cite{amos}    &  62.3  &  0.01   GeV$^2$ $<-t<$ 0.052 GeV$^2$      & 18  &  33  \\
 \cite{nagy}    &  62.1  &  0.825  GeV$^2$ $\le -t\le$ 2.25  GeV$^2$ & 27  & 167  \\
 \cite{star2}   & 200.0  &  0.045  GeV$^2$ $\le -t\le$ 0.135 GeV$^2$ & 35  & 120  \\
\hline
\multicolumn{3}{|c|}{\bf Total} & 222 & 858 \\
\hline
\end{tabular}
\end{center}
\vskip -0.2cm
\caption{The quality of the model description of the proton-proton EDS angular distributions.}
\label{tab2}
\end{table}

\begin{figure}[ht]
\epsfxsize=8.2cm\epsfysize=8.2cm\epsffile{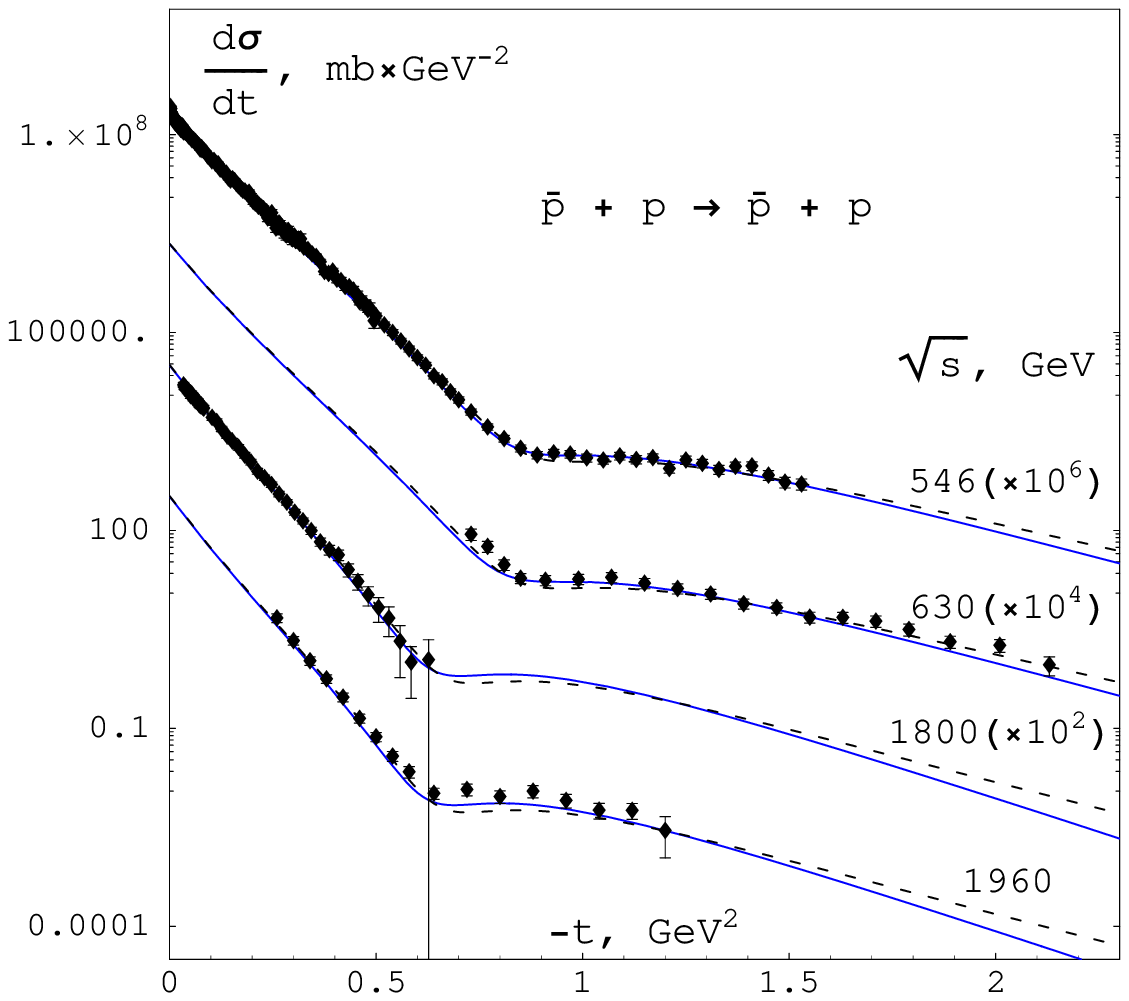}
\vskip -8.23cm
\hskip 8.8cm
\epsfxsize=8.2cm\epsfysize=8.2cm\epsffile{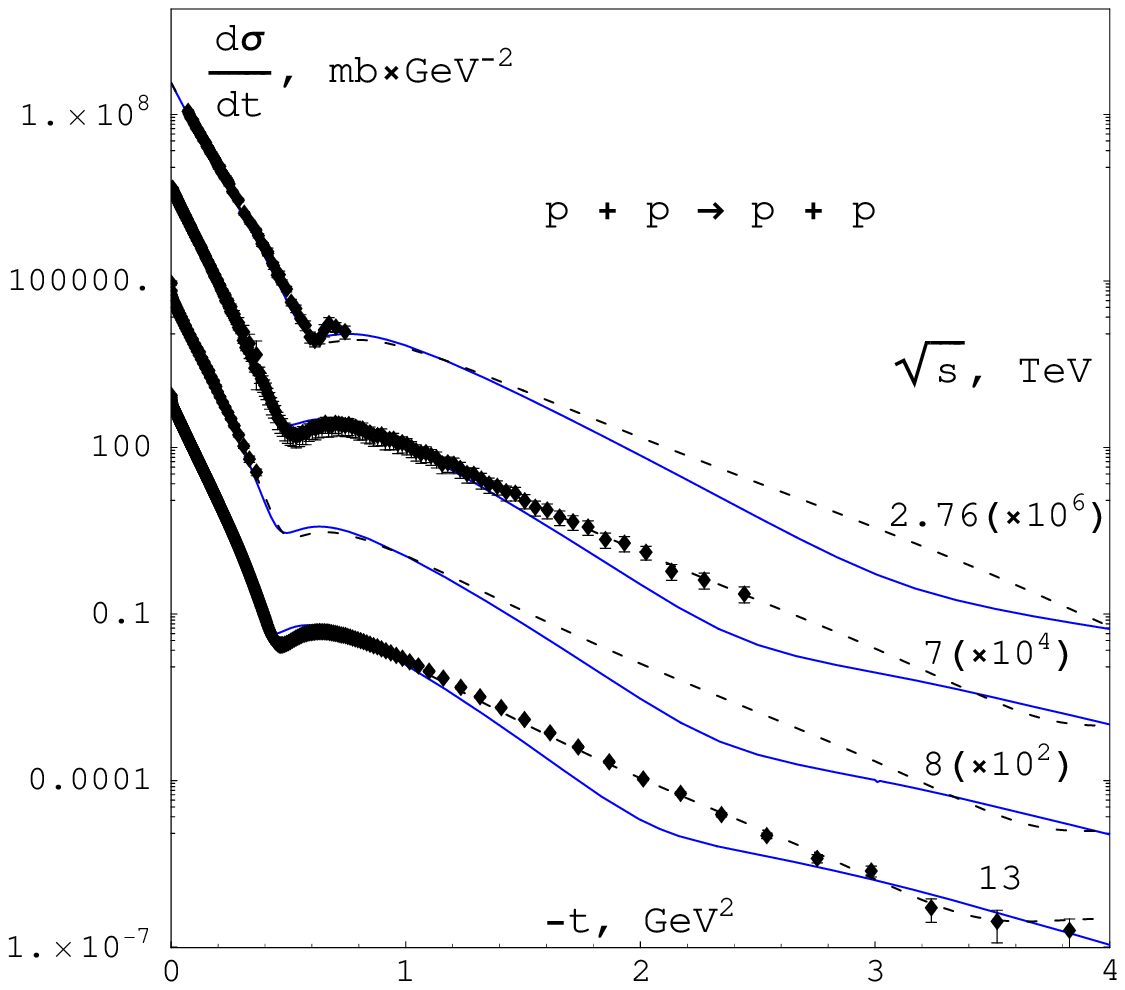}
\vskip -0.4cm
\caption{The differential cross-sections of nucleon-nucleon EDS at ultrahigh energies. The solid lines correspond to the two-Reggeon eikonal model. The dashed lines are related to the 
three-Reggeon approximation wherein the HP exchange contribution is taken into account.}
\label{high}
\end{figure}

\begin{figure}[ht]
\epsfxsize=8.2cm\epsfysize=8.2cm\epsffile{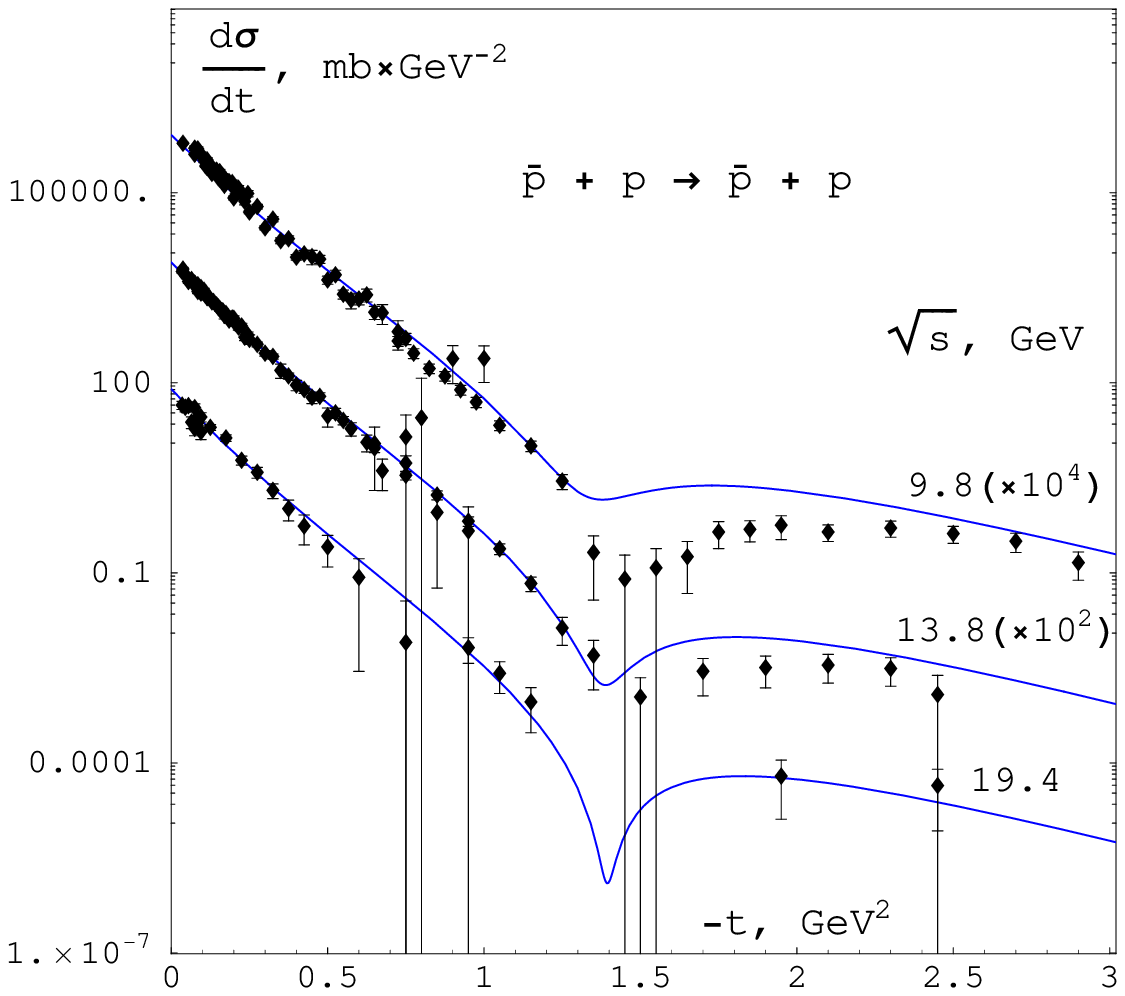}
\vskip -8.25cm
\hskip 8.8cm
\epsfxsize=8.2cm\epsfysize=8.2cm\epsffile{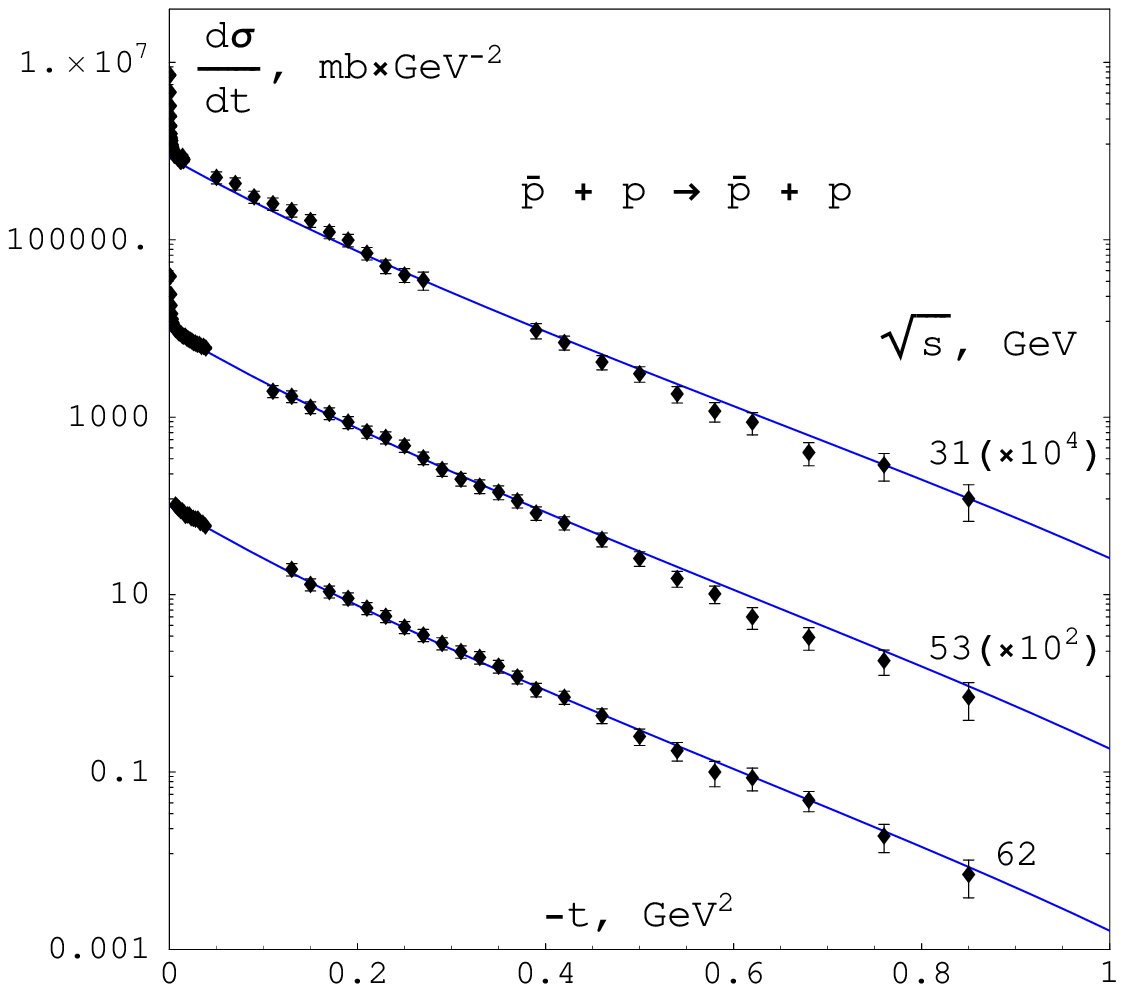}
\vskip -0.5cm
\caption{The $\bar p\,p$ EDS angular distributions.}
\label{applow}
\end{figure}

The predictions for the nucleon-nucleon differential cross-sections at ultrahigh energies\linebreak \cite{elaapp,elapp} and for the $\bar p\,p$ angular distributions at 
$\sqrt{s}\le$ 63 GeV \cite{amos,break0,cudell} are presented in Figs. \ref{high} (solid lines) and \ref{applow}, correspondingly. The quality of these predictions for the EDS data at 
$\sqrt{s}>$ 20 GeV is given in Tables \ref{tab3} and \ref{tab4}. 

The model predictions for the $p\,p$ total cross-section and for $\rho = \frac{{\rm Re}\,T_N(s,0)}{{\rm Im}\,T_N(s,0)}$ at $\sqrt{s}=$ 13 TeV are 
$\sigma^{model}_{tot}$(13 TeV) $\approx$ 109.6 mb and $\rho^{model}$(13 TeV) $\approx$ 0.129, while the corresponding measured values are 
$\sigma_{tot}$(13 TeV) = (110.5\,$\pm\,$2.4) mb and $\rho\,$(13 TeV) = $0.10\pm 0.01$ \cite{elapp}. It should be noted here that the extraction of these quantities (in particular, of 
the $\rho$-parameter) from the experimental angular distributions is a strongly model-dependent procedure.

Although the model demonstrates an impressive predictive value in the region of the SPS and Tevatron energies, one can 
observe significant deviations of the model curves from the data in the diffraction dip region at $\sqrt{s}<$ 30 GeV and $\sqrt{s}\ge$ 7 TeV and, also, in the region of relatively high 
transferred momenta, $\sqrt{-t}>$ 1.3 GeV. Let us explain these discrepancies.

\vskip -0.4cm

\newpage

\begin{table}[ht]
\begin{center}
\begin{tabular}{|l|l|l|l|l|}
\hline
\bf Experiment & $\sqrt{s}$, TeV & $t$-{\bf interval} & \bf Number of points & \bf $\chi^2$ \\
\hline
 UA4 ($\bar p\,p$)  &  0.546 & 0.01 GeV$^2$ $<-t<$ 0.035 GeV$^2$     &  50 &   43   (50) \\
 UA4 ($\bar p\,p$)  &  0.546 & 0.03 GeV$^2$ $<-t<$ 0.5 GeV$^2$       &  87 &  224  (143) \\
 UA4 ($\bar p\,p$)  &  0.546 & 0.46 GeV$^2$ $\le -t\le$ 1.53 GeV$^2$ &  34 &   36   (25) \\
 UA4 ($\bar p\,p$)  &  0.63  & 0.73 GeV$^2$ $\le -t\le$ 2.13 GeV$^2$ &  19 &   31   (27) \\
 E710 ($\bar p\,p$) &  1.8   & 0.03 GeV$^2$ $<-t<$ 0.63 GeV$^2$      &  51 &   20   (17) \\
 D0 ($\bar p\,p$)   &  1.96  & 0.26 GeV$^2$ $\le -t\le$ 1.2 GeV$^2$  &  17 &   31   (28) \\
 TOTEM ($p\,p$)     &  2.76  & 0.07 GeV$^2$ $<-t<$ 0.75 GeV$^2$      &  63 &  455  (268) \\
 TOTEM ($p\,p$)     &  7.0   & 0.01 GeV$^2$ $<-t<$ 2.5 GeV$^2$       & 161 &  206  (115) \\
 ATLAS ($p\,p$)     &  7.0   & 0.01 GeV$^2$ $<-t<$ 0.4 GeV$^2$       &  39 &   53   (52) \\
 TOTEM ($p\,p$)     &  8.0   & 0.01 GeV$^2$ $<-t<$ 0.2 GeV$^2$       &  22 &  143   (75) \\
 ATLAS ($p\,p$)     &  8.0   & 0.01 GeV$^2$ $<-t<$ 0.4 GeV$^2$       &  39 &  145  (193) \\
 TOTEM ($p\,p$)     & 13.0   & 0.01 GeV$^2$ $<-t<$ 3.9 GeV$^2$       & 403 & 9123 (2566) \\
\hline
\end{tabular}
\end{center}
\vskip -0.2cm
\caption{The quality of the model predictions for the nucleon-nucleon EDS angular distributions at ultrahigh energies. The numbers in parantheses are related to the three-Reggeon 
eikonal approximation wherein the HP exchange contribution is taken into account.}
\label{tab3}
\end{table}
\vskip -0.5cm

\begin{table}[ht]
\begin{center}
\begin{tabular}{|l|l|l|l|l|}
\hline
\bf Reference & $\sqrt{s}$, GeV & $t$-{\bf interval} & \bf Number of points & \bf $\chi^2$  \\
\hline
 \cite{nagy}    &  23.4 ($p\,p$)       &  0.825 GeV$^2$ $\le -t\le$ 2.25 GeV$^2$ & 27 & 217 \\
 \cite{break0}  &  31.0 ($p\,p$)       &  0.05  GeV$^2$ $\le -t\le$ 0.85 GeV$^2$ & 24 &  28 \\
 \cite{break0}  &  53.0 ($p\,p$)       &  0.11  GeV$^2$ $\le -t\le$ 0.85 GeV$^2$ & 24 &  21 \\
 \cite{break0}  &  62.0 ($p\,p$)       &  0.13  GeV$^2$ $\le -t\le$ 0.85 GeV$^2$ & 23 &  14 \\
 \cite{amos}    &  30.7 ($\bar p\,p$)  &  0.01  GeV$^2$ $<-t<$ 0.016 GeV$^2$     &  8 &  54 \\
 \cite{break0}  &  31.0 ($\bar p\,p$)  &  0.05  GeV$^2$ $\le -t\le$ 0.85 GeV$^2$ & 22 &  21 \\
 \cite{amos}    &  52.8 ($\bar p\,p$)  &  0.01  GeV$^2$ $<-t<$ 0.039 GeV$^2$     & 16 &  70 \\
 \cite{break0}  &  53.0 ($\bar p\,p$)  &  0.11  GeV$^2$ $\le -t\le$ 0.85 GeV$^2$ & 24 &  21 \\
 \cite{amos}    &  62.3 ($\bar p\,p$)  &  0.01  GeV$^2$ $<-t<$ 0.039 GeV$^2$     & 14 &  40 \\
 \cite{break0}  &  62.0 ($\bar p\,p$)  &  0.13  GeV$^2$ $\le -t\le$ 0.85 GeV$^2$ & 23 &   5 \\
\hline
\end{tabular}
\end{center}
\vskip -0.2cm
\caption{The quality of the model description of those ISR data on nucleon-nucleon EDS which were not included into the fitting procedure.}
\label{tab4}
\end{table}

\section*{5. Discussion of the issues}

\subsection*{5.1. Impact of other supercritical Reggeon contributions}

First of all, one can pay attention to the fact that the free parameter values related to the SP contribution are close to those obtained in the framework of the one-Reggeon 
approximation via fitting to the data on nucleon-nucleon EDS in the energy interval 546 GeV $\le\sqrt{s}\le$ 7 TeV \cite{godizov}.\linebreak Hence, the divergence between the model 
curves and the LHC data can be explained in the same way as it was done in \cite{godizov2}. Namely, the reason is our ignoring of the contribution into the eikonal by the hard Pomeron 
(HP), also known as the BFKL Pomeron. This supercritical Reggeon dominates in the deeply inelastic scattering (DIS) of leptons on protons. The intercept of its Regge trajectory can be 
extracted from the data on the proton unpolarized structure function $F^p_2(x,Q^2)$ \cite{struc} at high values of the incoming photon virtuality $Q^2$ and low values of the Bjorken 
scaling variable $x$: $\alpha_{\rm HP}(0)\approx 1.32$ \cite{godizov3}. In spite of such a high value of $\alpha_{\rm HP}(0)$, the corresponding term in the eikonal is subdominant 
relative to the SP contribution up to the LHC energies, due to the extremely weak $t$-dependence of $\alpha_{\rm HP}(t)$ in the transferred momentum range relevant for EDS (a detailed 
discussion of this matter can be found in \cite{godizov4}). 
\begin{figure}[ht]
\epsfxsize=8.2cm\epsfysize=8.2cm\epsffile{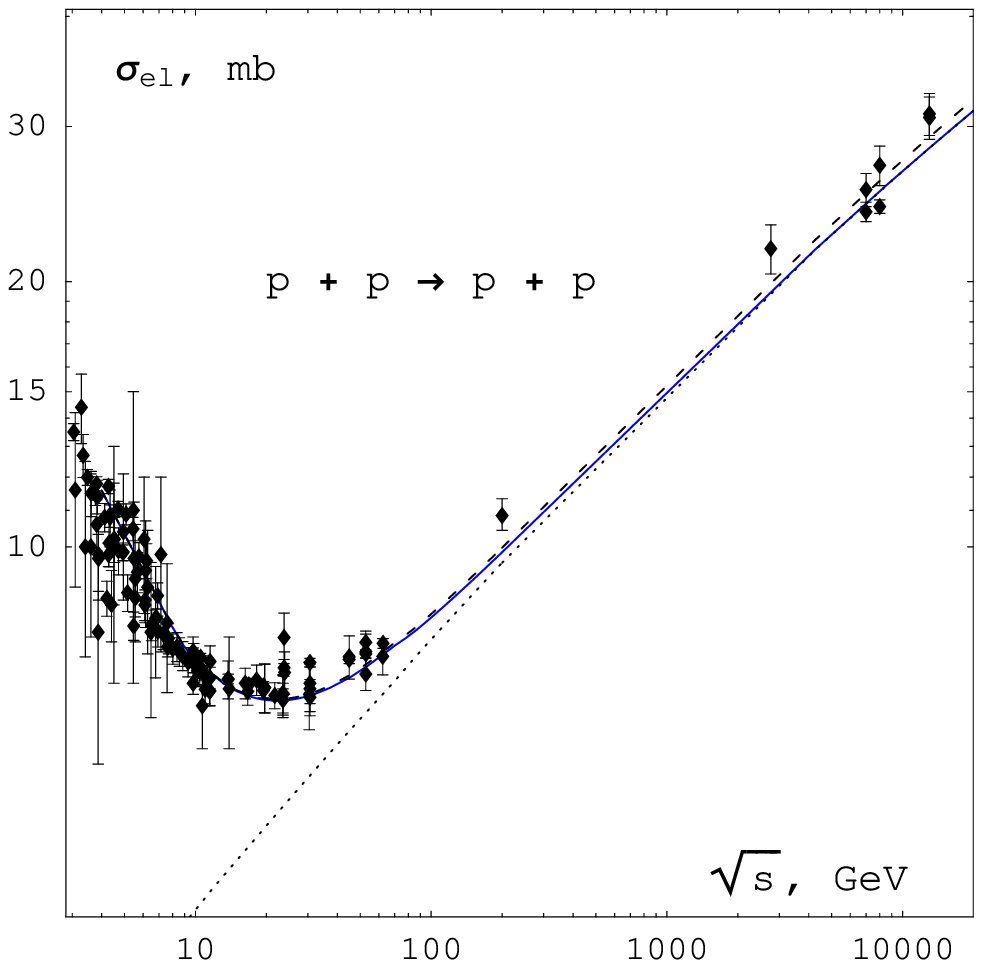}
\vskip -8.23cm
\hskip 8.8cm
\epsfxsize=8.2cm\epsfysize=8.2cm\epsffile{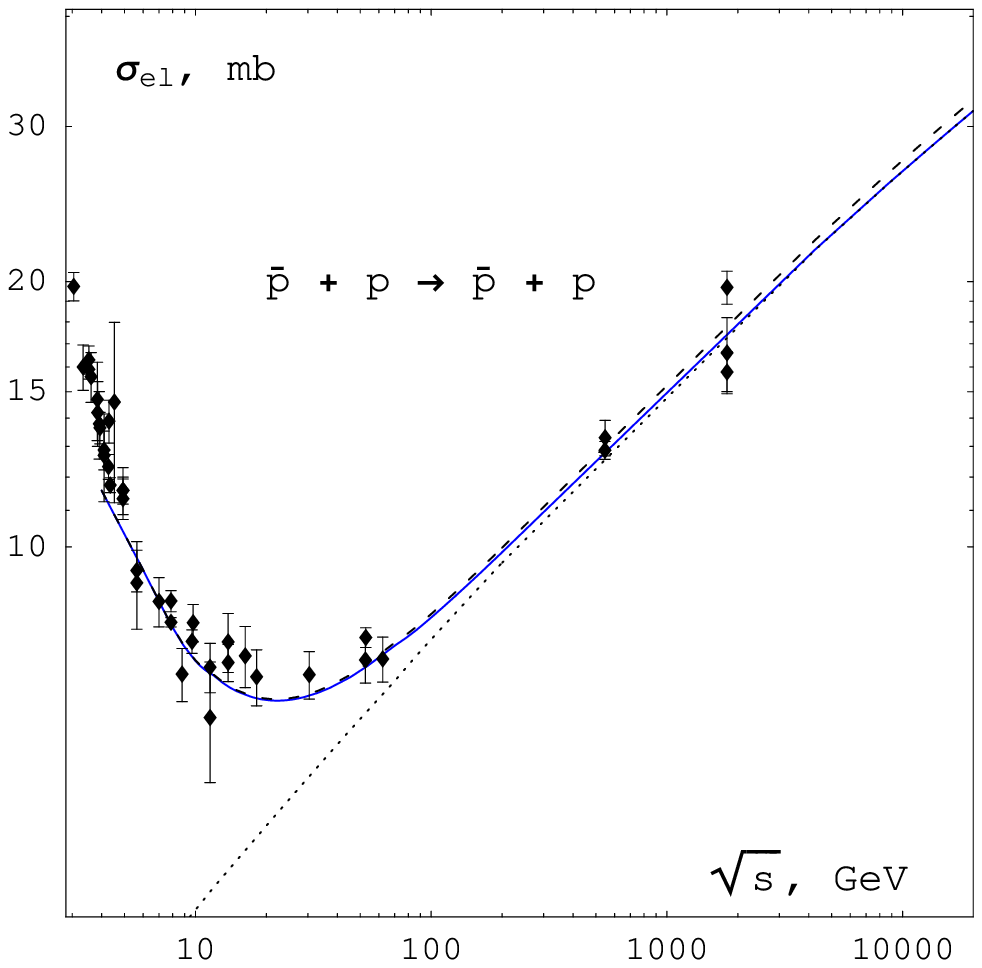}
\caption{The integrated cross-sections of nucleon-nucleon EDS. The solid lines correspond to the two-Reggeon eikonal model. The dashed lines are related to the three-Reggeon 
approximation, wherein the HP exchange contribution is taken into account. The dotted lines correspond to the one-Reggeon eikonal model, wherein just the SP exchanges are taken into 
account.}
\label{ela}
\end{figure}

If we add an extra term into the eikonal,
\begin{equation}
\label{eikhard}
\Omega_{\rm HP}(s,t)=\xi_+(\alpha_{\rm HP}(0))\,\beta^{(p)}_{\rm HP}(t)\left(\frac{s}{2s_0}\right)^{\alpha_{\rm HP}(0)}\,,
\end{equation}
where
\begin{equation}
\label{hardres}
\beta^{(p)}_{\rm HP}(t)\equiv g^{(p)2}_{\rm HP}(t)\,\alpha'_{\rm HP}(t) = \beta^{(p)}_{\rm HP}(0)\,e^{b\,t}
\end{equation}
is the HP Regge residue, $\beta^{(p)}_{\rm HP}(0)=0.0335$ and $b=1.6$ GeV$^{-2}$, then the deviations of the model curves from the LHC data become noticeably weaker (see Fig. 
\ref{high} (dashed lines) and Table \ref{tab3}). 

\begin{figure}[ht]
\vskip -0.1cm
\epsfxsize=8.2cm\epsfysize=8.2cm\epsffile{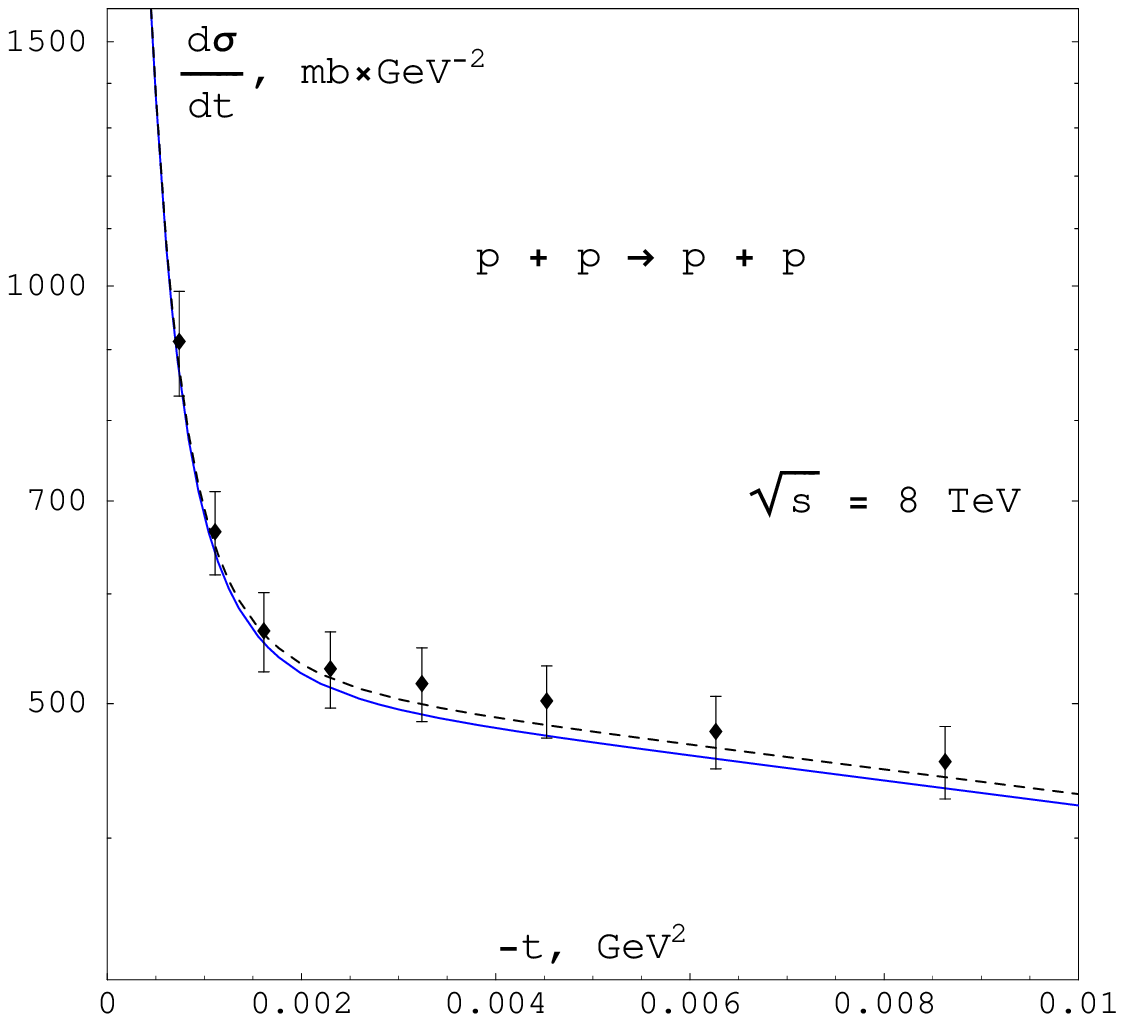}
\vskip -8.23cm
\hskip 8.8cm
\epsfxsize=8.2cm\epsfysize=8.2cm\epsffile{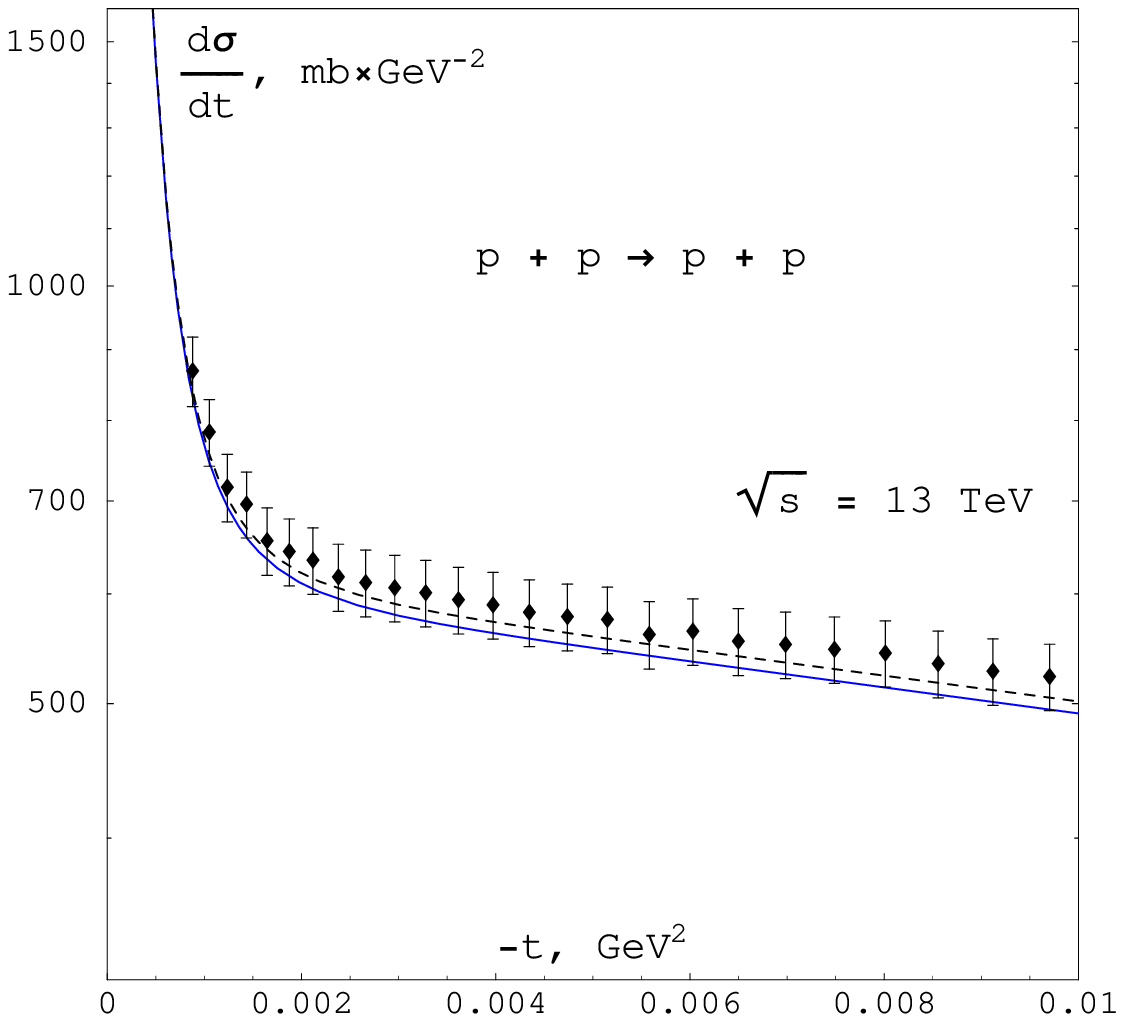}
\vskip -0.5cm
\caption{The differential cross-sections of proton-proton EDS in the CNI region. The solid lines correspond to the two-Reggeon eikonal model. The dashed lines are related to the 
three-Reggeon approximation, wherein the HP exchange contribution is taken into account.}
\label{totkul}
\end{figure}

However, the HP influence is crucial only if we consider the angular distributions in the dip region and at transferred momenta higher than 1 GeV. The HP exchanges have a negligible 
impact on the integrated cross-section values \cite{ppel,star2,appel} (Fig. \ref{ela}) and on the differential cross-section curves in the Coulomb-nuclear interference (CNI) region 
\cite{elapp} (Fig. \ref{totkul}). Here we take account of the electromagnetic interaction via direct adding of the Coulomb term to the full amplitude related to strong interaction: 
\begin{equation}
\label{elnuc2}
T_N(s,t)\to \Omega_C(s,t)+T_N(s,t)\,,\;\;\;\;\Omega_C(s,t) \approx \mp\frac{8\,\pi\,s\,\alpha_e}{t}F_E^2(t)\,,
\end{equation}
where $F_E(t)=\left(1-\frac{t}{0.71\,\rm GeV^2}\right)^{-2}$ is the dipole electric form-factor of proton, and $\alpha_e$ is the fine structure constant.\footnote{It was argued in 
\cite{godizov2} that, in the leading approximation, the CNI can be ignored in the amplitude level. As well, it was demonstrated in \cite{kaspar} on an example of two independent models 
that, at the LHC energies, such an approximation leads to a small (not higher than 3.5\%) overestimation of the differential cross-section. A more detailed discussion of the CNI 
phenomenon can be found in \cite{petrov}.}

Our ignoring of the HP exchanges in the framework of the two-Reggeon model is not the only reason for the model curve deviations from the data at the LHC energies. Some extra 
supercritical Reggeons should be included into consideration for the a better description, namely, the soft and hard Odderons (the $C$-odd counterpartners of the SP and HP, 
correspondingly). The exchanges by these Reggeons lead to a visible splitting of the $\bar p\,p$ and $p\,p$ angular distributions in the diffraction dip region \cite{odd}. 
Nonetheless, the Odderons' combined influence on the shape of the nucleon-nucleon EDS differential cross-section at ultrahigh energies seems to be a much finer effect than the impact 
of the HP exchanges, and, thus, it can be ignored, in the leading approximation.

\subsection*{5.2. Impact of secondary Reggeons}

\begin{figure}[ht]
\vskip -0.3cm
\epsfxsize=8.2cm\epsfysize=8.2cm\epsffile{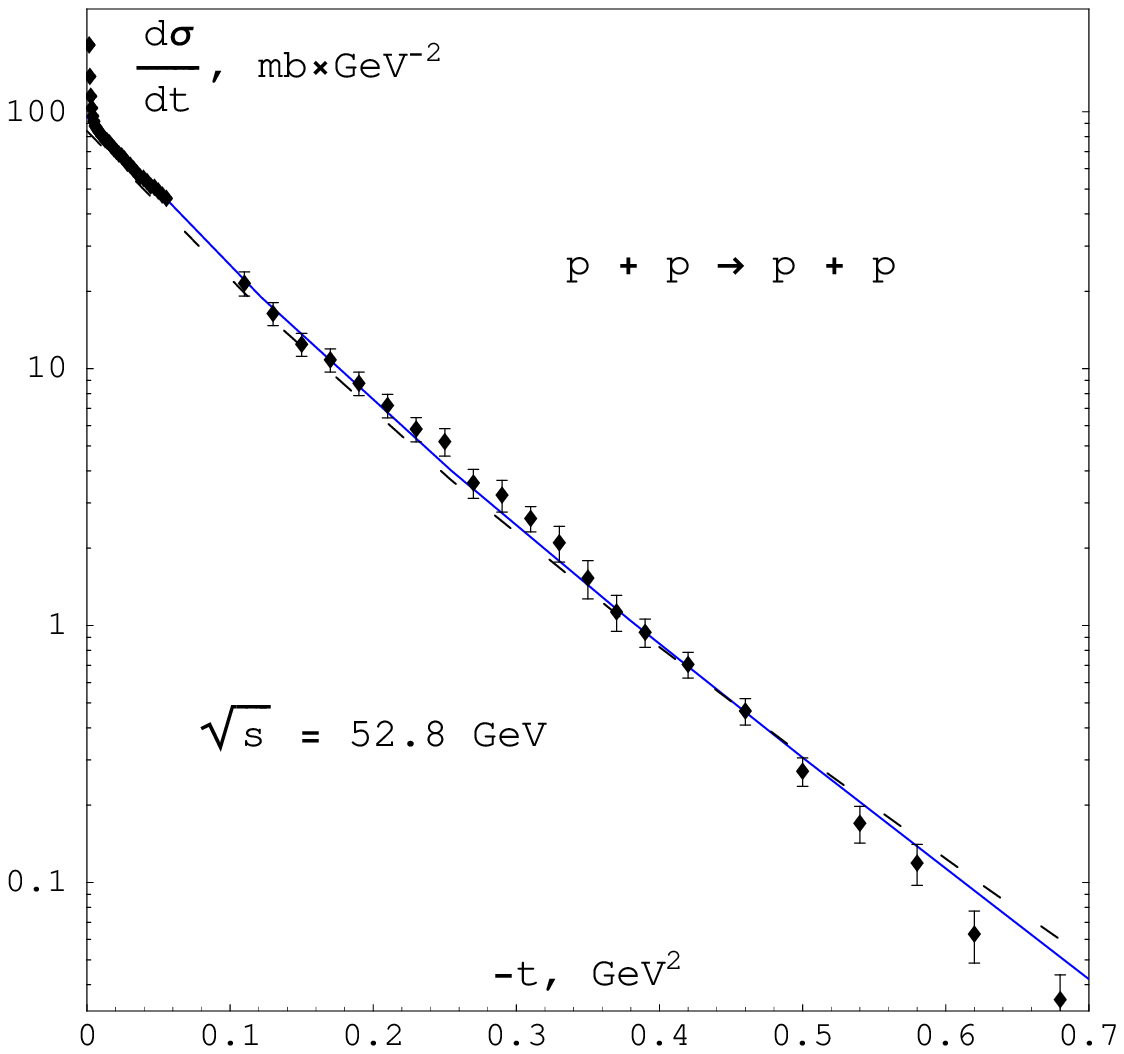}
\vskip -8.5cm
\hskip 8.25cm
\epsfxsize=8.75cm\epsfysize=8.75cm\epsffile{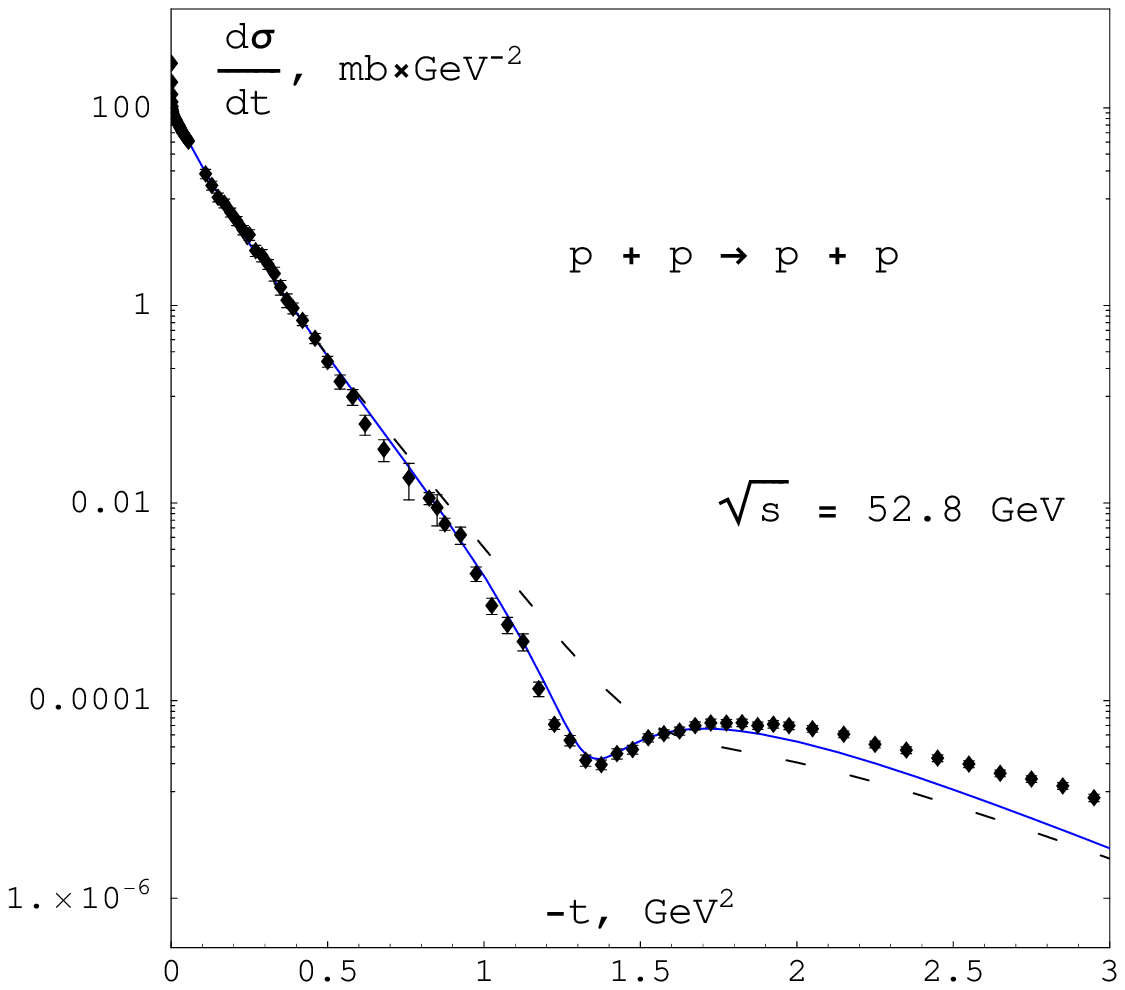}
\vskip -0.3cm
\caption{The $p\,p$ EDS angular distribution at $\sqrt{s}=$ 52.8 GeV. The solid lines correspond to the two-Reggeon eikonal model. The dashed lines are related to the 
one-Reggeon eikonal approximation wherein the FR exchange contribution to the eikonal is ignored.}
\label{pomer}
\end{figure}

To understand how strong is the influence of secondary Reggeons on the considered differential cross-sections, let us explore the impact of the FR exchanges on the proton-proton EDS 
angular distribution at $\sqrt{s}=$ 52.8 GeV. In Fig. \ref{pomer}, two curves are compared. The first one is obtained in the framework of the proposed two-Reggeon approximation (the 
solid line). The second one is related to the one-Reggeon approximation wherein the FR contribution into the eikonal is ignored (the dashed line), while the functions 
$\alpha_{\rm SP}(t)$ and $g^{(p)}_{\rm SP}(t)$ are kept the same. One can see that the impact of the FR exchanges on the differential cross-section shape is the most significant in 
the region of transferred momenta higher than 1 GeV. The distortion of the diffraction pattern at lower values of the transferred momentum is not so crucial. 

\begin{figure}[ht]
\vskip -0.6cm
\epsfxsize=8.2cm\epsfysize=8.2cm\epsffile{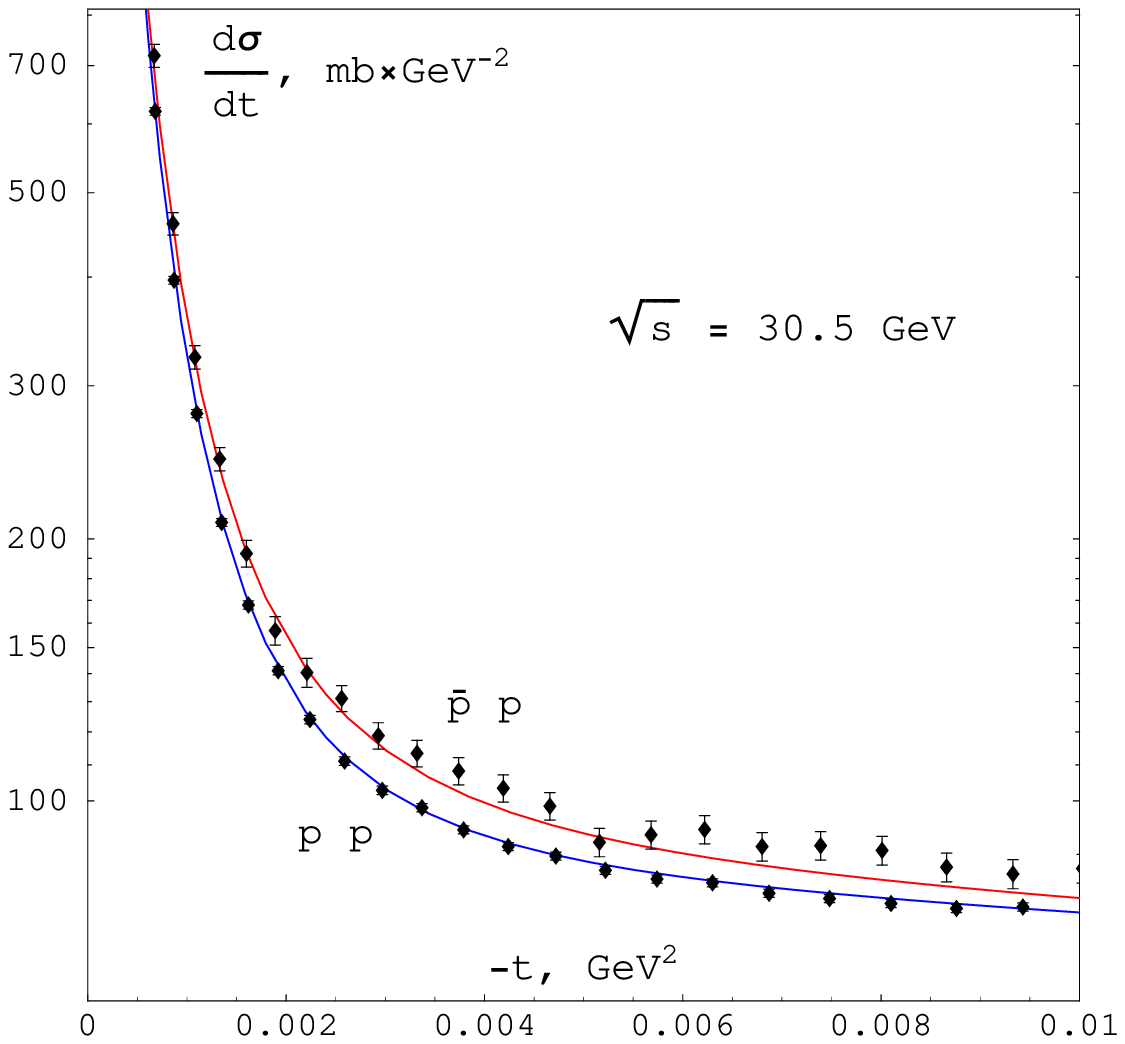}
\vskip -8.3cm
\hskip 8.8cm
\epsfxsize=8.18cm\epsfysize=8.18cm\epsffile{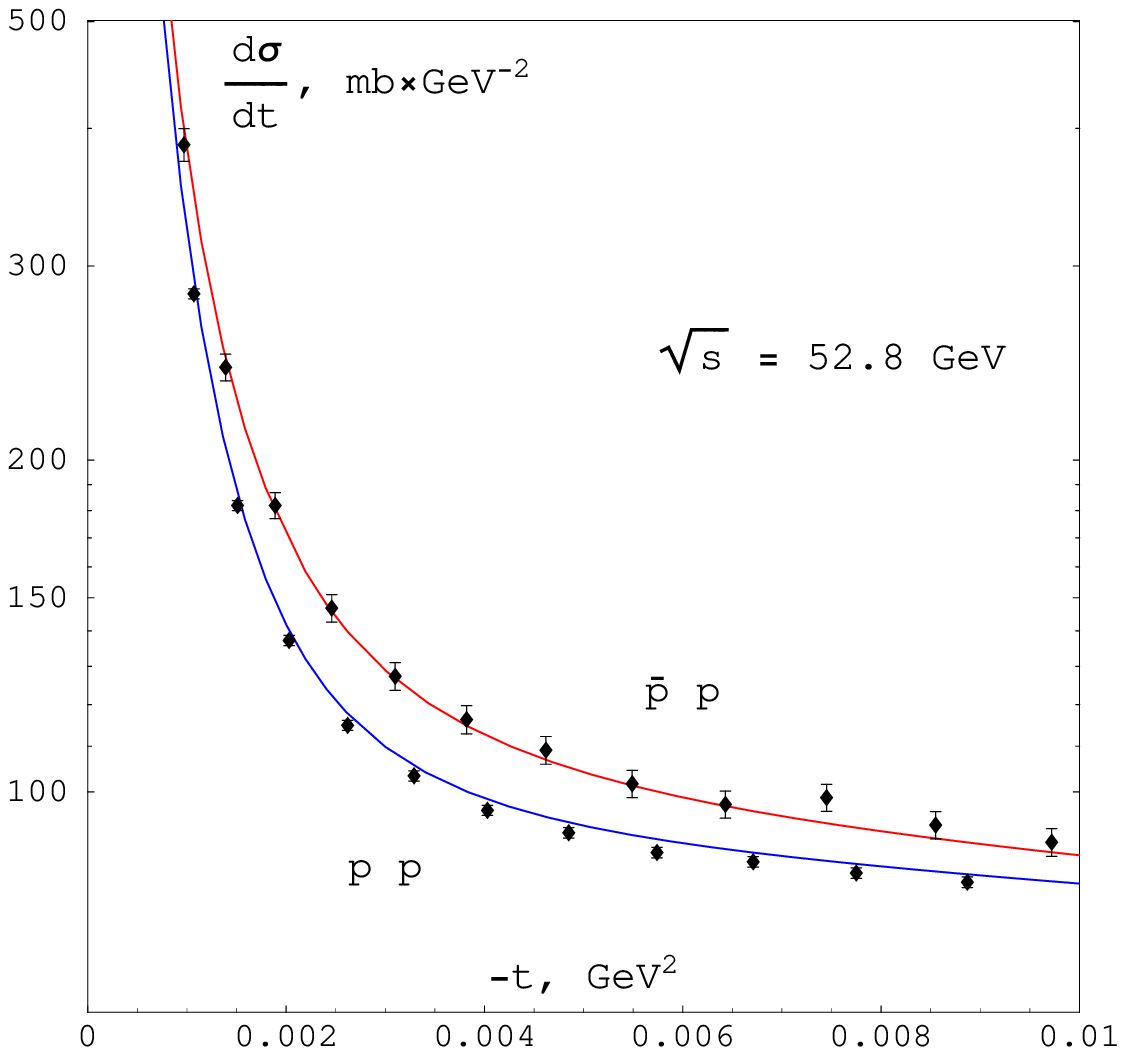}
\vskip -0.3cm
\caption{The differential cross-sections of the $p\,p$ and $\bar p\,p$ EDS at the ISR energies in the CNI region.}
\label{kul}
\end{figure}

Consequently, the observed deviations of the model curves from the data on the $\bar p\,p$ EDS at $\sqrt{s}<$ 15 GeV (see the left picture in Fig. \ref{applow}) can be explained by our 
ignoring of the combined influence of secondary Reggeons (namely, the $f$, $\omega$, $a$, and $\rho$ contributions of type 2\linebreak discussed above in subsection 3.1). The impact of 
secondaries goes down fast with energy and becomes almost negligible in the region $\sqrt{s}>$ 30 GeV. In its turn, the splitting between the $p\,p$ and $\bar p\,p$ angular 
distributions at the ISR energies in the CNI region \cite{amos} is, mainly, owing to electromagnetic interaction (see Fig. \ref{kul}). In other words, the combined contribution of 
secondary Reggeon exchanges into the eikonal is so subdominant that the two-Reggeon eikonal approximation turns out quite applicable for qualitative description of the $\bar p\,p$ EDS 
at $\sqrt{s}>$ 30 GeV.\linebreak

\subsection*{5.3. Other sources of divergences between the model curves and the experimental data}

The spin phenomena in the high-energy EDS of nucleons are negligible only at small enough values of the transferred momentum. Certainly, they should not be ignored in the region where 
the condition (\ref{suppr}) is invalid. Hence, the noticeable underestimation of the proton-proton EDS differential cross-sections at $\sqrt{s}<$ 63 GeV and $\sqrt{-t}>$ 1.3 GeV (see 
Fig. \ref{pplow}) can be related (at least, in part) to spin-flip effects.

Another reason of this underestimation is our choice of parametrizations for $\alpha_{\rm SP}(t)$, $g^{(p)}_{\rm SP}(t)$, $\alpha_{\rm FR}(t)$, and $g^{(p)}_{\rm FR}(t)$. The true SP 
and FR Regge trajectories and couplings to proton have much more complicated analytic behavior than the functions presented in (\ref{pomeron}). For instance, $g^{(p)}_{\rm SP}(t)$ and 
$g^{(p)}_{\rm FR}(t)$ even may be nonmonotonic at high enough values of the transferred momentum. Nonetheless, we restrict ourselves by the simplest parametrizations satisfying the 
relations (\ref{proper}), (\ref{proper2}), and (\ref{proper3}) -- (\ref{asypom2}) to make our reasoning more transparent and conclusions more justified. In other words, we have 
constructed the roughest model applicable for qualitative description of nucleon-nucleon EDS in a wide interval of the collision energy values. 

However, the simplicity of the proposed two-Reggeon approximation is not the only feature which distinguishes it from other models exploiting the notion of Reggeons.

\subsection*{5.4. Distinction from other Reggeon models}

Modern models for high-energy elastic scattering of hadrons based on Regge theory can be divided into two groups: Regge-eikonal models \cite{prokudin,selyugin2,khoze} and those models 
which take account of single and double Reggeon exchanges but ignore higher order terms in the eikonal power expansion of the scattering amplitude \cite{landshoff,pdg2014,martynov}.

The primary distinction of the proposed scheme from the above-mentioned models is the exploitation of nonlinear approximations to Regge trajectories in the region of 
negative values of the argument. A detailed discussion of why any hadron Regge trajectory must exhibit an essentially nonlinear behavior at $t<0$ and how such a nonlinearity may be 
consistent with the approximately linear behavior in the resonance region, $t>0$, can be found in \cite{petrov2}. Here we restrict ourselves just by one of the main arguments. 

Usage of linear parametrizations for Regge trajectories leads to the emergence of simple poles in the real part of the Reggeon signature factor (\ref{sign1}) at 
those negative values of $t$ where the corresponding Regge trajectory takes on integers 0, $-2$, $-4$, ... Various models resort to various tricks to avoid this issue. For instance, 
in \cite{prokudin}, the functions $\alpha(t)$ in the signature factors are replaced by $\alpha(0)$. Other models, as well, exploit some expressions for the even Reggeon signature 
factor which are different from (\ref{sign1}). However, that is not quite correct, because the Sommerfeld-Watson transform leads exactly to the formula (\ref{sign1}) \cite{collins}, 
and, hence, the expression (\ref{sign1}) is the only correct form of this factor. As a result, despite of its roughness and stiffness, the considered approximation exhibits a much 
higher predictive value than other Reggeon models, concerning the energy evolution of the diffraction pattern of nucleon-nucleon elastic scattering. 

Another feature of the proposed model is its above-discussed simplicity in the aspect of both the Reggeon structure of the scattering amplitude and the parametrizations of the Regge 
trajectories and residues. The vast majority of other models have a much more complicated Reggeon and parametric structure. For instance, three Pomerons are introduced within the 
model published in \cite{prokudin}, the relatively simple model considered in \cite{landshoff} includes the exchanges by the $\omega$-Reggeon with the trajectory significantly 
different from the Regge trajectory of the $f$-Reggeon, while the model proposed in \cite{martynov} introduces the so-called Froissaron and Maximal Odderon in addition to the usual 
simple $j$-pole Pomeron and Odderon and exploits a much higher number of free parameters as compared to other models. More complicated phenomenological structure makes these models 
more flexible and allows to include into consideration the kinematic region of transferred momenta higher than 1.5 GeV and, as well, to obtain a more satisfactory (from the 
statistical standpoint) description of available data, though the most of the above-mentioned models do not deal with the data in the collision energy region\linebreak 
$\sqrt{s}<$ 19 GeV.

\section*{6. Conclusions}

In view of the aforesaid, we have to conclude that the considered two-Reggeon eikonal model is a simple and reliable phenomenological tool which provides a qualitative description of 
the $p\,p$ EDS in the kinematic range \{10 GeV $\le\sqrt{s}\le$ 2 TeV, $\sqrt{-t}<$ 1.5 GeV\} and of the $\bar p\,p$ EDS in the kinematic range \{30 GeV $\le\sqrt{s}\le$ 2 TeV, 
$\sqrt{-t}<$ 1.5 GeV\}. The ultimate dominance of the SP and FR over other Reggeons in the EDS of nucleons up to the LHC energies is not just an artificial pattern but a real 
physical phenomenon. The obtained approximations to the SP and FR Regge trajectories and couplings to proton can be used in the framework of those Reggeon models which describe various 
reactions of high-energy inelastic diffractive scattering of nucleons, including single and double diffractive dissociation, central exclusive production of light vacuum resonance 
states, {\it etc.}

\end{document}